\documentclass[aps,prd,twocolumn,nofootinbib]{revtex4}
\pdfoutput=1
\usepackage{epsfig}
\usepackage{amsmath}
\usepackage{graphicx}
\usepackage{xcolor}
\usepackage{slashed}
\usepackage{multirow}
\usepackage{float}
\newcommand{\beq}{\begin{equation}}
\newcommand{\eeq}{\end{equation}}
\newcommand{\bqa}{\begin{eqnarray}}
\newcommand{\eqa}{\end{eqnarray}}

\newcommand{\eps}{\epsilon}

\newcommand{\als}{\alpha_s}

\begin{document}

%\begin{flushright}
%\today 
%\\  
%\end{flushright}

\title{Next-to-leading order QCD corrections to electromagnetic production and decay of fully charm tetraquarks }

\author{Xinran  Liu}
\affiliation{Department of Physics and Institute of Theoretical Physics, Nanjing Normal University, Nanjing,
Jiangsu 210023, China}
\affiliation{Nanjing Key Laboratory of Particle Physics and Astrophysics, Nanjing Normal University, Nanjing,
Jiangsu 210023, China}
\author{Yefan Wang}
\email{wangyefan@nnu.edu.cn}
\affiliation{Department of Physics and Institute of Theoretical Physics, Nanjing Normal University, Nanjing,
Jiangsu 210023, China}
\affiliation{Nanjing Key Laboratory of Particle Physics and Astrophysics, Nanjing Normal University, Nanjing,
Jiangsu 210023, China}
\author{Ruilin Zhu}
\email{rlzhu@njnu.edu.cn}
\affiliation{Department of Physics and Institute of Theoretical Physics, Nanjing Normal University, Nanjing,
Jiangsu 210023, China}
\affiliation{Nanjing Key Laboratory of Particle Physics and Astrophysics, Nanjing Normal University, Nanjing,
Jiangsu 210023, China}

\begin{abstract}
We investigate the electromagnetic properties of the fully charm tetraquark states, particularly incorporating contributions from internal gluon radiation. The paper first presents analytical expressions for the next-to-leading-order (NLO) QCD corrections to the decay amplitudes of fully charm tetraquarks into two photons. It is found that the QCD corrections are significant for the $J^{PC}=0^{++}$ and $J^{PC}=2^{++}$ fully charm tetraquark decay process. Subsequently, by considering photon-photon fusion in ultra-peripheral high-energy collisions of protons and nuclei and in electron-positron collision, we provide theoretical predictions for the production cross sections of fully charm tetraquark states. The results presented in this work regarding the electromagnetic production and decay of fully charm tetraquarks shall be tested in current and future experiments.
\end{abstract}

\maketitle

\section{Introduction}

It is well known that the standard quark model postulated by Gell-Mann and Zweig provided an excellent explanation for the hadron spectrum in the past century, such as the light meson nonet, light baryon octet, and decuplet, all of which are successful examples. However, in 2003, with the discovery of exotic hadronic states such as X(3872)~\cite{Choi:2003ue,Acosta:2003zx}, the situation changed. Up to now, a new-generation quark model that can neatly accommodate all these new hadronic states and explain all the production and decay properties has not yet been established.

In 2020, the LHCb Collaboration first reported a peak around
6900 MeV in the $J/\psi$-pair invariant mass distribution~\cite{LHCb:2020bwg}, which is considered the first candidate for a fully charm tetraquark state. Three years later, the 6900 MeV peak was confirmed by both the ATLAS and CMS Collaborations~\cite{ATLAS:2023bft,CMS:2023owd}. In the same year, a possible peak with mass $6267\pm43$ MeV and width $121 \pm72$ MeV was observed in the Belle experiment~\cite{Belle:2023gln}.
Very recently, the 6900 MeV peak and two other peaks at 6600 MeV and 7100 MeV in the $J/\psi$-pair channel and their spin-parity quantum numbers have been
determined by the CMS Collaboration using a data sample corresponding to an integrated luminosity of 135 $fb^{-1}$ at a center-of-mass energy of 13 TeV~\cite{CMS:2025fpt}.  Unlike the previous channel, the 6900 MeV peak has also been observed recently in the $J/\psi-\psi(2S)$ invariant mass distribution by  the ATLAS Collaboration~\cite{ATLAS:2025nsd}.

Correctly understanding the structure and properties of these exotic hadron states presents a significant theoretical challenge, as it involves the non-perturbative interactions of QCD that are difficult to handle. The pioneering idea of the possibility of four charm quark states near the $J/\psi$-pair threshold dates back to the 1970s-1980s \cite{Iwasaki:1975pv,Chao:1980dv}. In the new century, these exotic peaks in the $J/\psi$ pair spectrum have been studied using the tetraquark picture from the quark potential model~\cite{Anwar:2017toa,Karliner:2016zzc,Debastiani:2017msn,Wu:2016vtq,Liu:2019zuc,Jin:2020jfc,Liu:2021rtn,Faustov:2020qfm,Lu:2020cns,liu:2020eha,Wang:2019rdo}, diquark-antidiquark model~\cite{Bedolla:2019zwg,Zhu:2020xni,Giron:2020wpx}, 
gluonic tetraquark hybrid model~\cite{Tang:2024zvf, Tang:2024kmh, Tang:2025ept}, tetraquarks in QCD Sum Rules~\cite{Chen:2016jxd, Zhang:2020xtb, Wang:2020dlo, Wang:2020ols, Albuquerque:2020hio,Wang:2021mma,Wan:2020fsk,Yang:2020wkh,Wang:2017jtz, Wang:2018poa, Agaev:2023wua, Agaev:2023gaq}, tetraquarks in lattice QCD calculations~\cite{Li:2025vbd,Li:2025ftn,Meng:2024czd}, tetraquark-molecule mixing model~\cite{Santowsky:2021bhy},
tetraquarks in Bethe-Salpeter equations~\cite{Li:2021ygk, Ke:2021iyh,Heupel:2012ua}, dynamically generated resonance poles~\cite{Wang:2020wrp,Gong:2020bmg, Dong:2020nwy,Guo:2020pvt,Huang:2024jin}. The extension to other fully heavy tetraquark systems is studied in Refs.~\cite{Esposito:2018cwh,Agaev:2025wyf,Wang:2025apq,Xia:2025mgk}.  The production of fully charm tetraquarks is studied in hadron-hadron colliders~\cite{Feng:2023agq,Belov:2024qyi,Wang:2025hex,Celiberto:2025vra,Celiberto:2025ziy,Zhang:2020hoh,Feng:2020riv,Zhu:2020xni}.  Therein the first complete next-to-leading order  QCD calculation for the fully charm tetraquark states’ hadron production is given in Ref.~\cite{Wang:2025hex}. The quark and gluon to fully charm tetraquark fragmentation functions are studied in Refs.~\cite{Bai:2024flh,Bai:2024ezn,Celiberto:2024mab}. 
The decay properties of fully charm tetraquarks are studied in Refs.~\cite{Sang:2023ncm,Chen:2024orv,Becchi:2020uvq,Zhang:2023ffe,Wang:2023kir,Biloshytskyi:2022dmo,Chen:2022sbf,Lu:2025lyu}. A short review paper on the fully charm tetraquarks can be found in Ref.~\cite{Zhu:2024swp}.

In this paper, we investigate the interaction strength between photons and fully charm tetraquark states. Although their couplings are certainly weaker compared to processes involving gluons, such interactions can provide new observables that will contribute to a deeper understanding of exotic states in the $J/\psi$ pair spectrum. On the one hand, the decay of fully charm tetraquark states into two photons offers a new decay channel distinct from $J/\psi$ pair hadronic final states. Using photon final states allows for a clearer view of the internal structural features of fully charm tetraquark states. On the other hand, ultra-peripheral collisions (UPCs), which are reactions where two ions interact via their cloud of virtual photons, provide a novel production mechanism and offer new experimental detection opportunities. Related studies of exotic hadrons in UPCs can also be found in Refs.~\cite{Esposito:2021ptx,dEnterria:2025ecx}.

%\revise{In addition to the two-photon decay channel, the fully charm tetraquark can also decay into an electron-positron pair. Although the branching fraction is expected to be small, this channel also provides a clean signature in probing the electromagnetic structure of}

Since fully charm tetraquark states are composed entirely of heavy quarks, we can employ the nonrelativistic QCD (NRQCD) effective theory to separate the high- and low-energy scale interactions in their production and decay processes, thereby enabling factorized calculations for these processes~\cite{Bodwin:1994jh}.
The rest of this paper is organized as follows. In Sec.~\ref{sec:NRQCD}, we present the formula for the electromagnetic decays of fully charm tetraquarks. In Sec.~\ref{sec:NLO}, the next-to-leading order calculation of short-distance coefficients is obtained.   In Sec.~\ref{sec:UPC},  the $\gamma+\gamma\to T_{4c}$ photoproduction processes are studied. The cross sections of UPCs for different heavy-ion systems and electron-positron collisions are also given.   We will discuss the phenomenological numerical results in Sec.~\ref{sec:Res}.  The conclusion is given at the end.

\section{NRQCD factorization formula for fully charm tetraquark electromagnetic decays}\label{sec:NRQCD}
The NRQCD factorization formula for $T_{4c}\rightarrow\gamma\gamma$ at the lowest
order in velocity can be expressed as~\cite{Sang:2023ncm}  
\begin{align}
\Gamma(T^{0++}_{4c} \rightarrow \gamma\gamma) &= \frac{m_H}{128 m_c^4}\bigg( c_{1}\left\langle\mathcal{O}^{(0)}_{6\otimes\bar{6}}\right\rangle+c_{2}\left\langle\mathcal{O}^{(0)}_{\bar{3}\otimes3}\right\rangle
\nonumber \bigg.\\&+c_{\text{mix}}\left\langle\mathcal{O}^{(0)}_{\text{mix}}\right\rangle
\bigg),
\nonumber\\
\Gamma(T^{2++}_{4c} \rightarrow \gamma\gamma) &= \frac{m_H}{128 m_c^4} c_{3}\left\langle\mathcal{O}^{(2)}_{\bar{3}\otimes3}\right\rangle,
%\nonumber\\
%\Gamma(T^{2++}_{4c} \rightarrow e^+e^-) &= \frac{m_H}{128 m_c^4} c^{ee}_{3}\left\langle\mathcal{O}^{(2)}_{\bar{3}\otimes3}\right\rangle,
\label{eq:fac}
\end{align}
where the superscript in $T_{4c}$ denotes the spin-parity quantum number $J^{PC}$ of the $T_{4c}$, $m_c$ and $m_H$ are the masses of the charm quark and $T_{4c}$, respectively. The coefficients $\{c_i\}$ are the short distance coefficients (SDCs), which can be perturbatively calculated order by order. And the long distance matrix elements (LDMEs) are defined as 
\begin{align}
\left\langle\mathcal{O}^{(0)}_{6\otimes\bar{6}}\right\rangle &= \left|\left\langle0\left|\mathcal{O}^{(0)}_{6\otimes\bar{6}}\right|T_{4c}\right\rangle\right|^2,
\nonumber\\
\left\langle\mathcal{O}^{(J)}_{\bar{3}\otimes3}\right\rangle &= \left|\left\langle0\left|\mathcal{O}^{(J)}_{\bar{3}\otimes3}\right|T_{4c}\right\rangle\right|^2,
\nonumber\\
\left\langle\mathcal{O}^{(0)}_{mix}\right\rangle &= \text{Re}\left[\left\langle0\left|\mathcal{O}^{(0)}_{\bar{3}\otimes3}\right|T_{4c}\right\rangle \left\langle T_{4c}\left|\mathcal{O}^{(0)\dagger}_{\bar{6}\otimes6}\right|0\right\rangle \right],
\end{align}
where $J=0,2$. The above operators are constructed in the diquark-antidiquark configurations. In this assignment the S-wave spin-singlet diquark is a color-singlet while the S-wave spin-triplet diquark is a color-triplet~\cite{Zhu:2015bba,Wang:2017vnc,Zhu:2016arf}. Explicitly, these local operators read
\begin{align}
\mathcal{O}_{\mathbf{6} \otimes \overline{\mathbf{6}}}^{(0)}  =&\left[\psi_a^T\left(i \sigma^2\right) \psi_b\right]\left[\chi_c^{\dagger}\left(i \sigma^2\right) \chi_d^*\right] \mathcal{C}_{\mathbf{6} \otimes \overline{\mathbf{6}}}^{a b ; c d},\nonumber \\
\mathcal{O}_{\overline{\mathbf{3}} \otimes \mathbf{3}}^{(0)}  =&-\frac{1}{\sqrt{3}}\left[\psi_a^T\left(i \sigma^2\right) \sigma^i \psi_b\right]\left[\chi_c^{\dagger} \sigma^i\left(i \sigma^2\right) \chi_d^*\right] \mathcal{C}_{\overline{\mathbf{3}} \otimes \mathbf{3}}^{a b ; c d},\nonumber \\
\mathcal{O}_{\overline{\mathbf{3}} \otimes \mathbf{3}}^{(2)}  =&\left[\psi_a^T\left(i \sigma^2\right) \sigma^m \psi_b\right]\left[\chi_c^{\dagger} \sigma^n\left(i \sigma^2\right) \chi_d^*\right]\mathcal{C}_{\overline{\mathbf{3}} \otimes \mathbf{3}}^{a b ; c d}\nonumber\\ &\times\eps_H^{ij*} \Gamma^{i j ; m n},
\end{align}
where $\psi$ and $\chi$ are two dimensional Pauli spinor fields, $\sigma^i$ is the Pauli matrix, $\eps_H^{ij}$ is the polarization tensor of the $T_{4c}$. And the symmetric traceless color tensor is
\begin{align}
\Gamma^{i j ; m n} = \delta^{im}\delta^{jn} + \delta^{in}\delta^{jm} -\frac{2}{3}\delta^{ij}\delta^{mn}.
\end{align}
The color projection tensors are 
\begin{align}
\mathcal{C}_{\mathbf{6} \otimes \overline{\mathbf{6}}}^{ab;cd} &= \frac{1}{2\sqrt{6}}\left(\delta_{ac}\delta_{bd}+\delta_{ad}\delta_{bc}\right),\nonumber\\
\mathcal{C}_{\overline{\mathbf{3}} \otimes \mathbf{3}}^{ab;cd} &= \frac{1}{2\sqrt{3}}\left(\delta_{ac}\delta_{bd}-\delta_{ad}\delta_{bc}\right).    
\end{align}

\section{NLO QCD Corrections to SDCs}\label{sec:NLO}
With the factorization formula, the non-perturbative long-distance effects are absorbed into the LDMEs. Then the remaining SDCs $c_{i}$ can be perturbatively expanded in terms of the strong coupling constant $\als$,
\begin{align}
c_{i} = \sum_{j=0}^{\infty} \left(\frac{\als}{\pi}\right)^j c_{i,j} ,
\end{align}
To calculate the coefficient $c_{i,j}$, we first generate the amplitudes involving four free charm quarks $c\bar{c}c\bar{c}$, then project the four free charm quarks onto the targeted state with certain total angular momentum and parity. We use the spin operator $\Pi_{0}$ and $\Pi_{1}$ to obtain the spin-zero and spin-one diquarks, respectively. And the color projectors $\mathcal{C}_{\mathbf{6} \otimes \overline{\mathbf{6}}}^{ab;cd}$ and $\mathcal{C}_{\mathbf{3} \otimes \overline{\mathbf{3}}}^{ab;cd}$ are used to extract the color-sextet and color-antitriplet of the diquark, respectively. For the spin-singlet diquark contribution, the state of $c(p_1)\bar{c}(p_2)c(p_3)\bar{c}(p_4)$ is projected onto $J^{PC} = 0^{++}$ by the replacements
\begin{align}
&\bar{u}(p_1)X_1v(p_3)\bar{u}(p_2)X_2v(p_4)\rightarrow -\text{Tr}\left[\Pi_{0} X_2\Pi_{0} X_1^C\right]\mathcal{C}_{6\otimes\bar{6}}^{ab;cd},\nonumber\\
&\bar{u}(p_1)X_1v(p_4)\bar{u}(p_2)X_2v(p_3)\rightarrow \text{Tr}\left[\Pi_{0} X_2\Pi^{C}_{0} X_1^C\right]\mathcal{C}_{6\otimes\bar{6}}^{ab;cd},
\label{eq:rep1}
\end{align}
where the subscript $C$ stands for the charge conjugate, and the spin-singlet projector is~\cite{Qiao:2012hp,Qiao:2012vt} 
\begin{align}
\Pi_{0}= \frac{\left(\frac{\slashed{P}}{4}+m_c\right)\gamma_{5}}{\sqrt{2}}.
\end{align}
Here $P$ is the momentum of the $T_{4c}$, and the on-shell condition $P^2=m_H^2$ is satisfied. In the lowest order of velocity, the relative velocity between quarks is zero, then we have
\begin{align}
p_1=p_2=p_3=p_4=\frac{P}{4}.
\end{align}
The spin-triplet diquark contribution can be obtained by the replacements 
\begin{align}
&\bar{u}(p_1)X_1v(p_3)\bar{u}(p_2)X_2v(p_4)\nonumber\\\rightarrow& -\text{Tr}\left[\Pi_{1\mu} X_2\Pi_{1\nu} X_1^C\right]\mathcal{C}_{3\otimes\bar{3}}^{cd;ef}J^{\mu \nu}_{0,2},\nonumber\\
&\bar{u}(p_1)X_1v(p_4)\bar{u}(p_2)X_2v(p_3)\nonumber\\\rightarrow& \text{Tr}\left[\Pi_{1\mu} X_2\Pi^{C}_{1\nu} X_1^C\right]\mathcal{C}_{3\otimes\bar{3}}^{cd;ef}J^{\mu \nu}_{0,2},
\label{eq:rep2}
\end{align}
where the spin-triplet projector is
\begin{align}
\Pi_{1,\mu}= \frac{\left(\frac{\slashed{P}}{4}+m_c\right)\gamma_{\mu}}{\sqrt{2}}.
\end{align}
And the covariant projectors $J^{\mu \nu}_{0,2}$ in $D$ dimensions are
\begin{align}
J^{\mu \nu}_{0}&=\frac{1}{\sqrt{D-1}}\eta^{\mu \nu},\nonumber\\
J^{\mu \nu}_{2}&=\epsilon_{H,\alpha\beta}\left[\frac{1}{2}\eta^{\mu \alpha}\eta^{\nu \beta}+\frac{1}{2}\eta^{\mu \beta}\eta^{\nu \alpha}-\frac{1}{D-1}\eta^{\mu \nu}\eta^{\alpha\beta}\right],
\end{align}
where
\begin{align}
\eta^{\mu \nu} = -g^{\mu\nu} + \frac{P^\mu P^\nu}{m_H^2}.
\end{align}

\begin{figure}[ht]
	\centering
	\begin{minipage}{0.48\linewidth}
		\centering
		\includegraphics[width=0.8\linewidth]{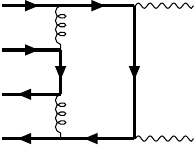}
	\end{minipage}
	\begin{minipage}{0.48\linewidth}
		\centering
		\includegraphics[width=0.8\linewidth]{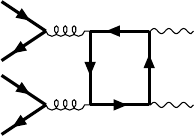}
	\end{minipage}
\caption{Typical one-loop Feynman diagrams for the $T_{4c}\rightarrow \gamma\gamma$. The thick black lines stand for the massive charm quarks.}
\label{OneLoop}
\end{figure}

Since there are no real emission processes in the NLO QCD corrections, the calculation remains within the framework of a two-body final state. After performing the trivial two-body phase space integration and the averaging of spin, the SDCs can be represented as
\begin{align}
c_{i} =     \frac{\left|\mathcal{M}_i\right|^2}{32\pi m_H(2J+1)},
\end{align}
where $J$ is the spin of $T_{4c}$.

To calculate the NLO corrections of $c_{i}$, we generate the tree-level and one-loop amplitudes of $T_{4c}\rightarrow\gamma\gamma$ by using the package {\tt FeynArts} \cite{Hahn:2000kx}. We find that there are 40 tree and 856 one-loop Feynman diagrams contributing to $T_{4c}\rightarrow\gamma\gamma$. The typical Feynman diagrams are plotted in Figure \ref{OneLoop}. Since the tree-level amplitudes are easy to calculate, here we describe the computing procedure of the one-loop amplitudes. After substituting the fermion chains with eq. (\ref{eq:rep1}) and eq. (\ref{eq:rep2}), the package {\tt FeynCalc} \cite{Shtabovenko:2020gxv,Shtabovenko:2023idz} is used to simplify the Dirac matrices. Then the one-loop bare amplitudes can be projected into several Lorentz structures $\{F_i\}$,
\begin{align}
\mathcal{M}^{\text{bare}} = \sum_{i} d_iF_i.
\end{align}
In the one-loop amplitudes, the corresponding coefficients $\{d_i\}$ are the linear combinations of one-loop scalar integrals, which can be reduced to a set of basis integrals called master integrals (MIs) due to the identities from integration by parts (IBP) \cite{Tkachov:1981wb,Chetyrkin:1981qh} with the package {\tt Kira} \cite{Klappert:2020nbg}. Then the analytical results of one-loop MIs can be easily obtained by package {\tt FeynHelpers} \cite{Shtabovenko:2016whf} based on {\tt Package-X} \cite{Patel:2015tea}. 

The one-loop amplitudes computed above contain UV divergences, which will be canceled with the diagrams from counterterms, arising from the renormalization of the $\als$, charm quark masses and quark-field strength. The charm quark mass and field are renormalized in the on-shell scheme while the strong coupling $\als$ is renormalized in the $\overline{\rm MS}$ scheme,
\begin{align}
m_{c,bare} &= Z_m m_c,\nonumber\\
\psi_{c,bare} & = \sqrt{Z_2}\psi_{c},\nonumber\\
\als^{bare}C^{\eps} &= \mu^{2\eps} \als Z_{\als},    
\end{align}
where $ C_{\eps} = (4\pi e^{- \gamma_E} )^{\eps} $ within D-dimension $D=4-2\eps$. And the renormalization constants are
\begin{align}
Z_{\als} &= 1+ \left(\frac{\als}{4\pi}\right)\frac{2n_l-31}{3\eps}+\mathcal{O}(\als^2),\nonumber\\
Z_m &= 1 +\left(\frac{\als}{4\pi}\right)D_{\eps}\left(-\frac{4}{\epsilon }-\frac{16}{3}-\frac{32 \epsilon }{3}\right)+\mathcal{O}(\als^2),\nonumber\\
Z_2&= 1 +\left(\frac{\als}{4\pi}\right)D_{\eps}\left(-\frac{4}{\epsilon }-\frac{16}{3}-\frac{32 \epsilon }{3}\right)+\mathcal{O}(\als^2),
\end{align}
where $C_A = 3$ and $C_F = 4/3$ have been substituted, and we have
\begin{align}
     D_{\eps} \equiv \frac{ \Gamma(1+\eps)}{e^{-\gamma_E\eps}} \left(\frac{ \mu^2}{m_c^2}\right)^{\eps}.
\end{align}

The results of SDCs at LO read \cite{Sang:2023ncm}  
\begin{align}
c_{1,0} &= \frac{128\pi^3e_c^4\alpha^2\als^2}{3m_Hm_c^4},\nonumber\\
c_{2,0} &= \frac{2304\pi^3e_c^4\alpha^2\als^2}{m_Hm_c^4},\nonumber\\
c_{mix,0} &= \frac{256\sqrt{6}\pi^3e_c^4\alpha^2\als^2}{m_Hm_c^4},\nonumber\\
c_{3,0} &= \frac{16384\pi^3e_c^4\alpha^2\als^2}{15m_Hm_c^4},
\end{align}
And the NLO corrections can be expressed as
\begin{align}
c_{i,1} = c_{i,0}  \left(\frac{\als}{\pi}\right) X_{i,1},
\end{align}
where $X_{i,1}$ is the dimensionless ratio. 
Note that the amplitude square in $c_{mix,1}$ is defined as
\begin{align}
2\text{Re}\left[\mathcal{M}^{\text{ren}}_{1,1}\mathcal{M}^{\dagger}_{2,0}+\mathcal{M}^{\text{ren}}_{2,1}\mathcal{M}^{\dagger}_{1,0}\right].
\end{align}

The analytic results for spin-0 tetraquark electromagnetic decay are as follows
\begin{widetext}
\begin{align}
X_{1,1} =& \frac{\left(31-2n_l\right)}{6}\log \left(\frac{\mu^2}{m_c^2}\right)+\frac{5}{32} \log ^2\left(7-4 \sqrt{3}\right)
-\frac{151}{48}\log ^2\left(2 \sqrt{2}+3\right)
+\frac{559}{18 \sqrt{2}}\log \left(2 \sqrt{2}+3\right)
+\frac{673}{72} \text{Li}_2\left(-\frac{1}{3}\right)\nonumber\\
&-\frac{527}{144}\text{Li}_2\left(\frac{1}{3}\right)
+\frac{91}{96} \left(-4 \text{Li}_2\left(\frac{1}{5}\right)-2 \text{Li}_2\left(\frac{3}{5}\right)+2 \text{Li}_2\left(-\frac{1}{5}\right)+4 \text{Li}_2\left(-\frac{3}{5}\right)+3 \log ^2(3)-2 \log (5) \log (3)\right)
\nonumber\\&
+\frac{207-16n_l}{64} \pi ^2-\frac{1595-36n_l}{54}\log(2)-\frac{21-8n_l}{18}
\nonumber\\
\approx &\left(5.16667-0.333333n_l\right)\log \left(\frac{\mu^2}{m_c^2}\right)+31.6078-1.56086n_l,
\nonumber\\
X_{2,1} =&\frac{\left(31-2n_l\right)}{6}\log \left(\frac{\mu^2}{m_c^2}\right)-\frac{11}{288} \log ^2\left(7-4 \sqrt{3}\right)
-\frac{5}{16}\log ^2\left(2 \sqrt{2}+3\right)+\frac{139 \log ^2(3)}{288}
+\frac{673}{648} \text{Li}_2\left(-\frac{1}{3}\right)
\nonumber\\&
-\frac{95}{1296}\text{Li}_2\left(\frac{1}{3}\right)+\frac{7}{144} \left(2 \text{Li}_2\left(\frac{1}{5}\right)+ \text{Li}_2\left(\frac{3}{5}\right)- \text{Li}_2\left(-\frac{1}{5}\right)-2 \text{Li}_2\left(-\frac{3}{5}\right)+\log (3)\log (5) \right)
\nonumber\\&
+\frac{31-16n_l}{576} \pi ^2
-\frac{401}{162 \sqrt{2}}\log \left(2 \sqrt{2}+3\right)
-\frac{67-36n_l}{54}\log(2)+\frac{65-24n_l}{54}
\nonumber\\
\approx &\left(5.16667-0.333333n_l\right)\log \left(\frac{\mu^2}{m_c^2}\right)-3.00931-0.256502n_l,
\nonumber\\
X_{mix,1}&=\frac{\left(31-2n_l\right)}{6}\log \left(\frac{\mu^2}{m_c^2}\right)
+\frac{17}{288} \log ^2\left(7-4 \sqrt{3}\right)
-\frac{83}{48}\log ^2\left(2 \sqrt{2}+3\right)+\frac{479 \log ^2(3)}{288}
+\frac{3365}{648} \text{Li}_2\left(-\frac{1}{3}\right)
\nonumber\\&
-\frac{2419}{1296}\text{Li}_2\left(\frac{1}{3}\right)+\frac{133}{144} \left(-2 \text{Li}_2\left(\frac{1}{5}\right)-\text{Li}_2\left(\frac{3}{5}\right)+\text{Li}_2\left(-\frac{1}{5}\right)+2 \text{Li}_2\left(-\frac{3}{5}\right)-\log (3)\log (5) \right)
\nonumber\\&
+\frac{947-80n_l}{576} \pi ^2+\frac{2315}{162 \sqrt{2}}\log \left(2 \sqrt{2}+3\right)-\frac{277-12n_l}{18}\log(2)+\frac{1}{54},
\nonumber\\
\approx &\left(5.16667-0.333333n_l\right)\log \left(\frac{\mu^2}{m_c^2}\right)+14.2992-0.908680n_l.
\end{align}

The analytic result for spin-2 tetraquark electromagnetic decay is
\begin{align}
X_{3,1} =&\frac{\left(31-2n_l\right)}{6}\log \left(\frac{\mu^2}{m_c^2}\right)
+\frac{479}{144} \text{Li}_2\left(-\frac{1}{3}\right)-\frac{2431}{1152}\text{Li}_2\left(\frac{1}{3}\right)
+\frac{21}{128} \log ^2\left(7-4 \sqrt{3}\right)-\frac{615}{256}\log ^2\left(2 \sqrt{2}+3\right)
\nonumber\\&
+\frac{467}{768} \log ^2(3)-\frac{2665 \pi ^2}{4608}-\frac{49}{16 \sqrt{3}}\log \left(4 \sqrt{3}+7\right)
+\frac{575}{96 \sqrt{2}}\log \left(2 \sqrt{2}+3\right)
-\frac{22357}{2592}\log (2)+\frac{463-168n_l}{432}
\nonumber\\
\approx &\left(5.16667-0.333333n_l\right)\log \left(\frac{\mu^2}{m_c^2}\right)-15.1997-0.388889n_l,
\end{align}
where $n_l$ is the number of the light quark flavor. 
\end{widetext}

\section{Application to the photoproduction of fully charm tetraquarks}\label{sec:UPC}

We have obtained explicit NLO QCD corrections to the electromagnetic decays of fully charm S-wave tetraquarks in the preceding section. Next we employ them to obtain the NLO QCD corrections to the photoproduction of fully charm tetraquarks.  The parton level amplitude in the $\gamma+\gamma\to T_{4c}$ photoproduction process is conveniently obtained from the parton level amplitude  in the $ T_{4c}\to \gamma+\gamma $ decay process through the crossing symmetry.

The NRQCD factorization formula for the $\gamma+\gamma\to T_{4c}$ photoproduction process  at the lowest
order in velocity can be written as  
\begin{align}
\sigma(\gamma\gamma\rightarrow T^{0++}_{4c} ) = &\frac{\pi^2\delta(1-z)}{32m_H^2 m_c^4}\bigg( c_{1}\langle0|O^{0}_{6\otimes\bar{6},6\otimes\bar{6}}|0\rangle\nonumber\\&+c_{2}\langle0|O^{0}_{\bar{3}\otimes3,\bar{3}\otimes3}|0\rangle
\nonumber \bigg.\\&+c_{\text{mix}}\langle0|O^{0}_{\bar{3}\otimes3,6\otimes\bar{6}}|0\rangle
\bigg),
\nonumber\\
\sigma(\gamma\gamma\rightarrow T^{2++}_{4c} )  = &\frac{5\pi^2\delta(1-z)}{32m_H^2 m_c^4} c_{3}\langle0|O^{2}_{\bar{3}\otimes3,\bar{3}\otimes3}|0\rangle,
\label{eq:fac2}
\end{align}
where $z=m_H^2/\hat{s}$ with $\hat{s}$ as the center-of-mass energy of two photons.  $O^{J}_{g_1,g_2}$ with $g_i=6\otimes\bar{6}$ or $\bar{3}\otimes3$ are the production operators for
tetraquarks with spin quantum number $J$
\begin{align}
O_{g_1,g_2}^{J}=\mathcal{O}_{g_1}^{(J)} \sum_X\left|T_{4 c}^J+X\right\rangle\left\langle T_{4 c}^J+X\right| \mathcal{O}_{g_2}^{(J) \dagger},
\end{align}
where the NRQCD operators $\mathcal{O}_{g_1}^{(J)}$ have been defined in the previous section. By employing the vacuum saturation effect, a connection can be established between the NRQCD production LDMEs and the decay LDMEs. 

Then we can apply our formulae to the production via photon fusion in high-energy ultraperipheral 
collisions of protons and nuclei. The UPC cross section via photon fusion can be written as 
\begin{align}
\sigma(\mathrm{A}\; \mathrm{B}\,\xrightarrow{\gamma\gamma} \mathrm{A}  \; \mathrm{B}\; T_{4c})=
\int \frac{dE_{\gamma_1}}{E_{\gamma_1}} \frac{dE_{\gamma_2}}{E_{\gamma_2}} \, \frac{\mathrm{d}^2\sigma^{(\mathrm{AB})}_{\gamma_1,\gamma_2}}{\mathrm{d}E_{\gamma_1}\mathrm{d}E_{\gamma_2}}\,,
\label{eq:UPCs}
\end{align}
where  $E_{\gamma_i}$ is the photon energy and the values of maximum energy $E_{\gamma_i}^\mathrm{max}$ for different UPCs at the LHC can be found in Ref.~\cite{Shao:2022cly}. In general, the photon number densities in UPCs are dependent 
on both the photon energy and the impact parameters from hadrons A and B. If we ignore the hadronic-nonoverlap effects, the two photon cross section
distribution becomes~\cite{Shao:2022cly,Jiang:2025pad}
\begin{align}
\frac{\mathrm{d}^2\sigma^{(\mathrm{AB})}_{\gamma_1,\gamma_2}}{\mathrm{d}E_{\gamma_1}\mathrm{d}E_{\gamma_2}} =  n_{\gamma/\mathrm{A}}(\bar{x}_1 )n_{\gamma/\mathrm{B}}(\bar{x}_2)\sigma(\gamma\gamma\rightarrow T_{4c} )\,,\label{eq:UPCs2}
\end{align}
where $\bar{x}_i=x_im_N b_{min}$ and $x_i=E_{\gamma_i}/E_{beam}$.
$m_N=0.93$ GeV is the nucleon mass. The minimum of impact parameter $b_{min}$ is set to be the nuclear radius as  $b_{min}=R_A\approx 1.2 A^{\frac{1}{3}}fm$ with the ion mass number $A$. The effective photon distribution function is 
\begin{align}
n_{\gamma/\mathrm{A}}(\chi)=&\frac{2Z^2\alpha}{\pi}\left[\left(1-\gamma_\mathrm{L}^{-2}\right)\frac{\chi^2}{2}\left(K_0^2(\chi)-K_1^2(\chi)\right)\right.
\nonumber\\&\left.+\chi K_0(\chi)K_1(\chi)\right]\,,
\end{align}
where $Z$ is the charge number and $\gamma_\mathrm{L}$ is the Lorentz boost factor for ion beams. $K_{0}$ and $K_{1}$ are modified Bessel functions.

In a similar way, one can expect the fully charm tetraquark
production via photon fusion in electron-positron collisions. The production cross section via two-photon fusion in electron-positron collisions can be written as~\cite{ParticleDataGroup:2024cfk}
\begin{align}
\sigma(e^-e^+\xrightarrow{\gamma\gamma} e^-e^+ T_{4c}) =& (2J + 1) \frac{8\alpha^2 \, \Gamma_{\gamma\gamma}}{m_T^3}\left[- \frac{1}{3} \left( \ln \frac{s}{m_R^2} \right)^3  
\right.\nonumber\\&\left.+
    f\left(\frac{m_T^2}{s}\right) 
    \left( \ln \frac{m_V^2 s}{m_e^2 m_T^2} - 1 \right)^2   
\right],
\end{align}
where $\alpha $ is the electromagnetic coupling and $s$ is the collision center of mass energy. $m_V$ is usually chosen as the $\rho$ meson mass. The auxiliary function is defined as 
\begin{align}
f(z)=& (1+\frac{z}{2})^2 \log(\frac{1}{z})-\frac{(1-z)(3+z)}{2}.
\end{align}

\section{Discussion and Numerical Results}\label{sec:Res}
With the analytical expressions of NLO corrections at hand, we can obtain the numerical results of SDCs up to NLO corrections. Firstly we take the input parameters from \cite{ParticleDataGroup:2024cfk,CMS:2023owd} 
\begin{align}
&m_H = 4m_c = 6847^{+44+48}_{-28-20} \,\text{MeV},\quad m_b = 5 \,\text{GeV},
\nonumber\\&
m_w =   91.1876 \,\text{GeV},
\quad
\als(m_w) = 0.1181, \quad n_l=3,
\end{align}
where the fully charm tetraquark mass is chosen from the latest CMS measurement.
Then the package {\tt{RunDec}} \cite{Herren:2017osy} was used to obtain $\als$ at other scales. The strong coupling $\als$ at $2m_c$, $4m_c$ and $8m_c$ is evaluated to be 
\begin{align}
&\als(2m_c) = 0.242997,\quad    \als(4m_c) = 0.195959,\nonumber\\& \als(8m_c) = 0.166230.
\end{align}
To perform the decay width, we still need the LDMEs as in Eq. (\ref{eq:fac}). In this work, we take the LDMEs in Ref.~\cite{Wang:2025hex}, 
in which we have
\begin{align}
\langle\mathcal{O}_{\bar{3}\otimes3}^{(2)}\rangle\mathcal{B}(T_{4c}) =4.44^{+1.60+2.58+1.24}_{-1.60-1.14-0.00}\times10^{-5}\text{GeV}^9,
\end{align}
where $\mathcal{B}(T_{4c})$ stands for the branch ratio of $T_{4c}\rightarrow2J/\psi$.
The first column in uncertainties is from the existing LHCb measurements; the second column is from
the scale uncertainty with
and the third column is from the parton distribution
function sets. It should be noted that, due to the difference between the production and decay processes, the LDMEs used in the decay calculation are related to those extracted from production by a factor of $1/(2J+1)=1/5$ from the Vacuum Saturation Approximation~\cite{Bodwin:1994jh}. In the absence of experimental data for 0++ fully charm tetraquark, we assume that 
\begin{align}
&\quad\langle\mathcal{O}_{\bar{6}\otimes6}^{(0)}\rangle =  \langle\mathcal{O}_{\bar{3}\otimes3}^{(0)}\rangle =  \langle\mathcal{O}_{mix}^{(0)}\rangle
\nonumber\\
&=4.44^{+1.60+2.58+1.24}_{-1.60-1.14-0.00}\times\frac{10^{-5}}{\mathcal{B}(T_{4c})}\text{GeV}^9.
\end{align}
This equality corresponds to mix angle $\theta=\pi/4$ in \cite{Zhang:2023ffe}. For simplicity, here we define
\begin{align}
\mathcal{R}_{\gamma\gamma}(T_{4c})\equiv\Gamma(T_{4c}\rightarrow\gamma\gamma)\mathcal{B}(T_{4c}\rightarrow2J/\psi).
\end{align}

\begin{figure}[ht]
	\centering
	\includegraphics[width=1\linewidth]{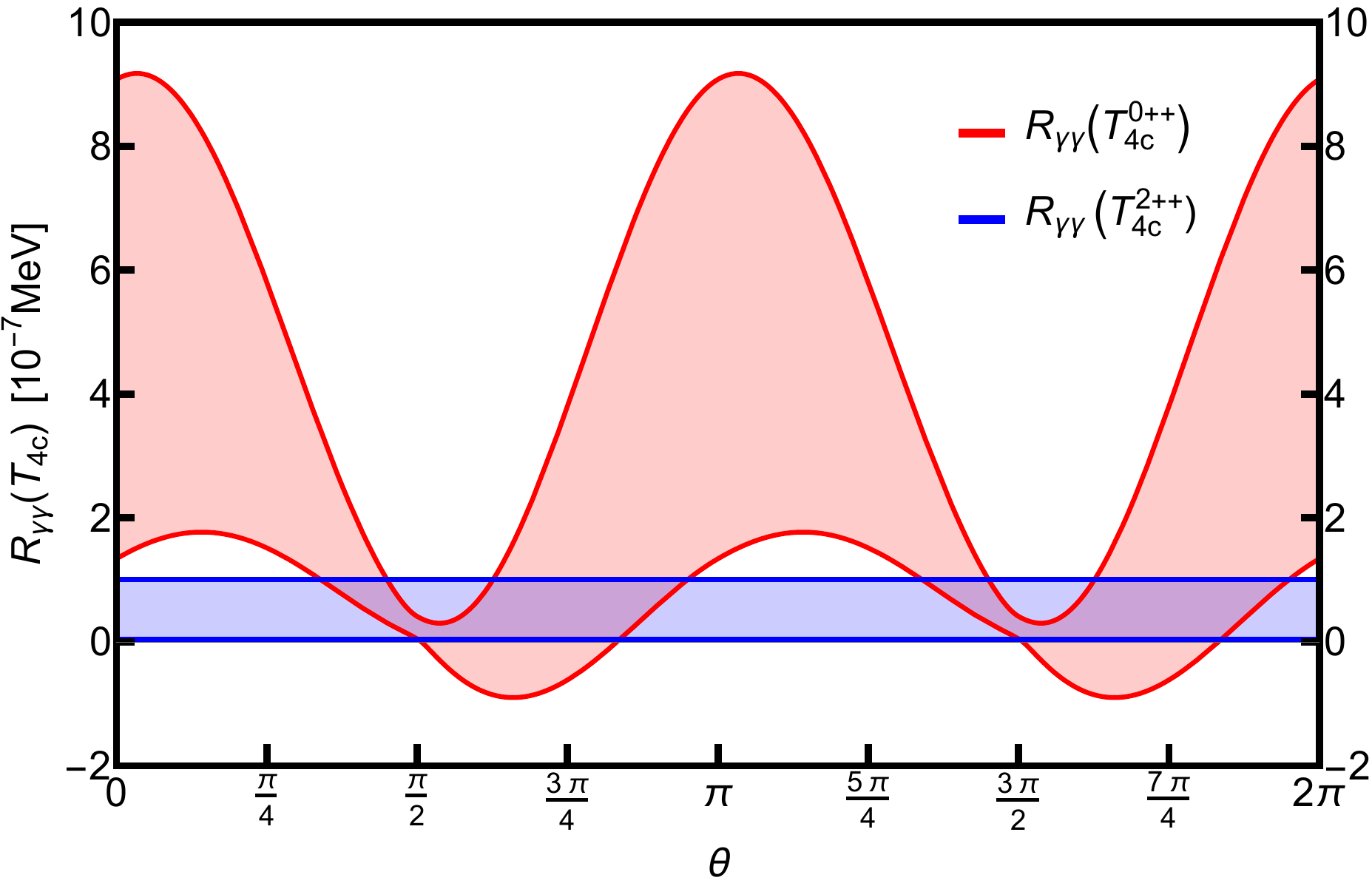}
\caption{The numerical values of $\mathcal{R}^{\theta}_{\gamma\gamma}(T^{0++}_{4c})$ and $\mathcal{R}_{\gamma\gamma}(T^{2++}_{4c})$ with the different mix angle $\theta$. The error bars denote the uncertainties.}
\label{plottheta}
\end{figure}

\begin{table*}[]
    \centering
    \caption{The numerical values of  $\mathcal{R}_{\gamma\gamma}(T^{0++}_{4c})$ and $\mathcal{R}_{\gamma\gamma}(T^{2++}_{4c})$ with different scales.}
    \label{tab:num}
    \scalebox{0.95}{
        \begin{tabular}{|c|c|c|c|c|}
            \hline
            & [$10^{-8}$ MeV] & $\mu = 2m_c$ & $\mu = 4m_c$ & $\mu = 8m_c$ \\
            \hline
            \multirow{2}{*}{$\mathcal{R}_{\gamma\gamma}(T^{0++}_{6\otimes\bar{6},4c})$}
            & LO  & 0.39 & 0.25 & 0.18 \\
            \cline{2-5}
            & NLO  & 1.37 & 0.86 & 0.60 \\
            \cline{2-5}
            \hline
            \multirow{2}{*}{$\mathcal{R}_{\gamma\gamma}(T^{0++}_{\bar{3}\otimes3,4c})$} 
            & LO   & 20.89 & 13.59 & 9.78 \\
            \cline{2-5}
            \cline{2-5}
            & NLO  & 24.12 & 20.17 & 16.79 \\
            \cline{2-5}
            \hline
            \multirow{2}{*}{$\mathcal{R}_{\gamma\gamma}(T^{0++}_{mix,4c})$} 
            & LO   & 5.69 & 3.70 & 2.66 \\
            \cline{2-5}
            \cline{2-5}
            & NLO  & 13.32 & 9.03 & 6.73 \\
            \cline{2-5}
            \hline
            \multirow{2}{*}{$\mathcal{R}_{\gamma\gamma}(T^{0++}_{4c})$}
            & LO   &  26.97&  17.54&  12.62\\
            \cline{2-5}
            \cline{2-5}
            & NLO  &  38.81&  30.06&  24.12\\
            \cline{2-5}
            \hline
            \multirow{2}{*}{$\mathcal{R}_{\gamma\gamma}(T^{2++}_{4c})$}
            & LO   & 9.90 & 6.44 & 4.63 \\
            \cline{2-5}
            \cline{2-5}
            & NLO  & 1.79 & 4.51 & 4.87 \\
            \cline{2-5}
            \hline
        \end{tabular}
    }
\end{table*}
\begin{figure}[ht]
	\centering
	\includegraphics[width=1\linewidth]{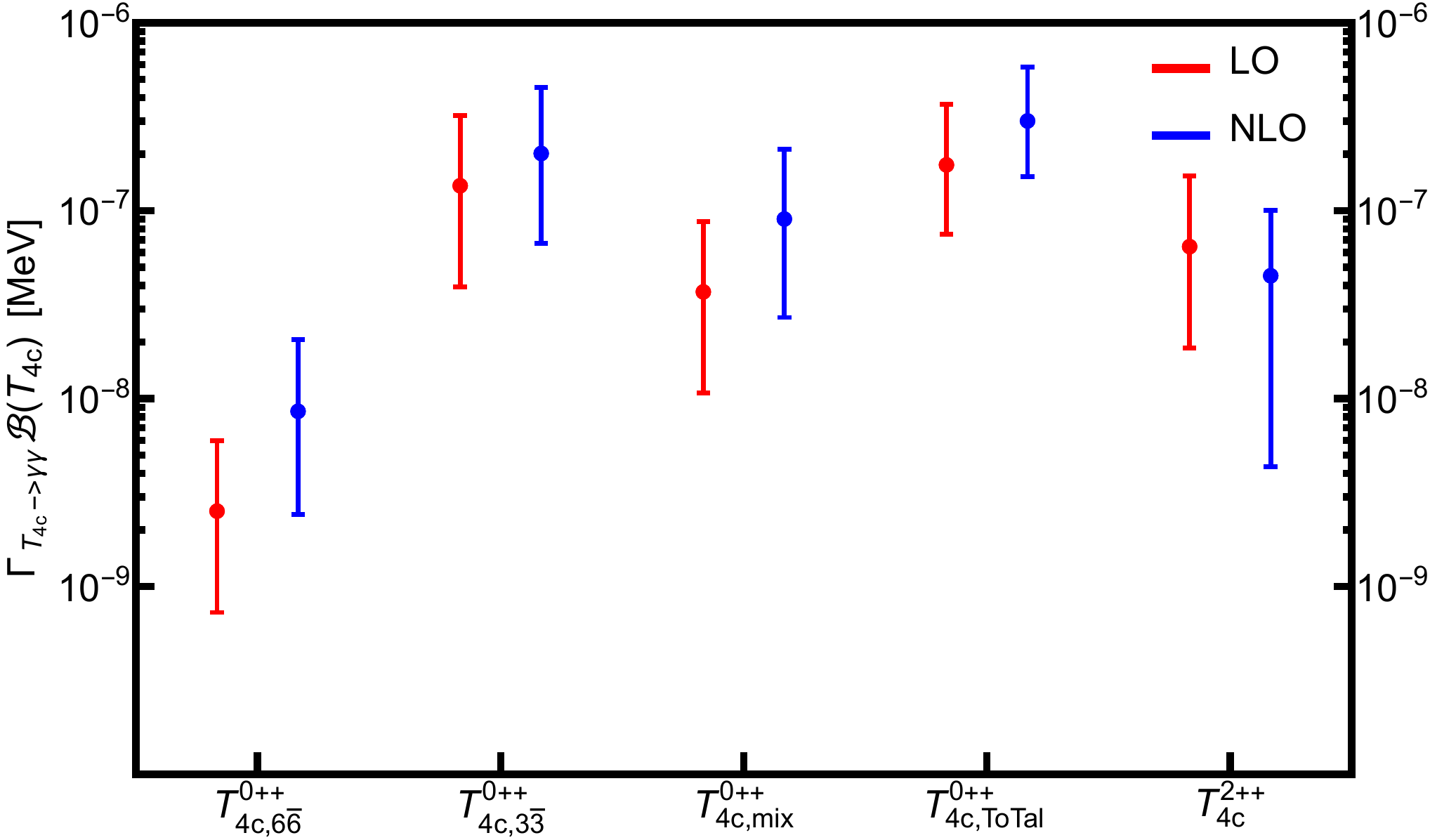}
\caption{The numerical values of $\mathcal{R}_{\gamma\gamma}(T^{0++}_{4c})$ and $\mathcal{R}_{\gamma\gamma}(T^{2++}_{4c})$. The error bars denote the uncertainties.}
\label{plotscheme}
\end{figure}
In Table \ref{tab:num}, we show the numerical results of $\mathcal{R}_{\gamma\gamma}(T_{4c})$ up to NLO with different renormalization scale $\mu$. All other inputs, including the LDMEs and $m_c$, are set to their central values. Here we have 
$\mathcal{R}_{\gamma\gamma}(T^{0++}_{4c}) = \mathcal{R}_{\gamma\gamma}(T^{0++}_{6\otimes\bar{6},4c})+\mathcal{R}_{\gamma\gamma}(T^{0++}_{\bar{3}\otimes3,4c})+\mathcal{R}_{\gamma\gamma}(T^{0++}_{mix,4c})$. Among the three components of $T^{0++}_{4c}$, $R_{\gamma\gamma}(T^{0++}_{6\otimes\bar{6},4c})$ are quite smaller than others, $R_{\gamma\gamma}(T^{0++}_{\bar{3}\otimes3,4c})$ gives the dominant contribution. Then $\mathcal{R}_{\gamma\gamma}(T_{4c})$ of $T^{0++}_{4c}$ is much larger than  $T^{2++}_{4c}$. At LO, $\mathcal{R}_{\gamma\gamma}(T^{0++}_{4c})$ is about 3 times that of $\mathcal{R}_{\gamma\gamma}(T^{2++}_{4c})$ at the typical scale $\mu=4m_c$. And one can see that the QCD corrections are quite large. The NLO QCD corrections increase the LO value of $\mathcal{R}_{\gamma\gamma}(T^{0++}_{4c})$ by approximately 70\% at $\mu=4m_c$. In contrast, the correction for the $2++$ state is negative, decreasing the LO prediction by about 30\% at $\mu=4m_c$. After considering the all uncertainties including $m_c$ and LDMEs, up to NLO we have
\begin{align}
\mathcal{R}_{\gamma\gamma}(T^{0++}_{6\otimes\bar{6},4c}) &= 0.86^{+1.20}_{-0.61}\times10^{-8}\text{MeV},\nonumber\\
\mathcal{R}_{\gamma\gamma}(T^{0++}_{\bar{3}\otimes3,4c}) &= 20.17^{+25.22}_{-13.47}\times10^{-8}\text{MeV},\nonumber\\
\mathcal{R}_{\gamma\gamma}(T^{0++}_{Mix,4c}) &= 9.03^{+12.14}_{-6.33}\times10^{-8}\text{MeV},\nonumber\\
\mathcal{R}_{\gamma\gamma}(T^{0++}_{4c}) &= 30.00^{+28.01}_{-14.89}\times10^{-8}\text{MeV},\nonumber\\
\mathcal{R}_{\gamma\gamma}(T^{2++}_{4c}) &= 4.51^{+5.54}_{-4.07}\times10^{-8}\text{MeV}.
\end{align}
To show these properties explicitly, we plot the results in Figure \ref{plotscheme}. We also consider the dependence of the mixing angle $\theta$ in \cite{Zhang:2023ffe}, in which the $\mathcal{R}_{\gamma\gamma}(T^{0++}_{4c})$ can be expressed as
$\mathcal{R}^{\theta}_{\gamma\gamma}(T^{0++}_{4c}) = 2\sin^2(\theta)\mathcal{R}_{\gamma\gamma}(T^{0++}_{6\otimes\bar{6},4c})+2\cos^2(\theta)\mathcal{R}_{\gamma\gamma}(T^{0++}_{\bar{3}\otimes3,4c})
+2\sin(\theta)\cos(\theta)\mathcal{R}_{\gamma\gamma}(T^{0++}_{mix,4c})$. The numerical results for $\mathcal{R}^{\theta}_{\gamma\gamma}(T^{0^{++}}_{4c})$ as a function of $\theta$ are shown in Figure \ref{plottheta}, together with the $\mathcal{R}_{\gamma\gamma}(T^{2^{++}}_{4c})$ value for comparison.

\begin{table}[h]
	\centering
    	\caption{Photon fusion cross sections of $T^{0++}_{4c}$ and $T^{2++}_{4c}$ in various UPCs processes $(\mathrm{A}\; \mathrm{B}\,\xrightarrow{\gamma\gamma} \mathrm{A}  \; \mathrm{B}\; T_{4c})$ with $\theta=\pi/4$ and ${\mathcal{B}(T_{4c})}=0.5$}.
	\label{tab:cs}
	\scalebox{0.95}{
		\begin{tabular}{|c|c|c|c|}
			\hline
            System & $E^i_{beam}$ (TeV) & $\sigma(T^{0++}_{4c})$ & $\sigma(T^{2++}_{4c})$ \\
            \hline
            {P-P} & {7.0+7.0 TeV} & $0.13^{+0.12}_{-0.07}$ fb & $0.10^{+0.12}_{-0.09}$ fb \\
            \hline
            {O-O} & {3.5+3.5 TeV}  & $0.16^{+0.15}_{-0.08}$ pb & $0.12^{+0.15}_{-0.11}$ pb \\
            \hline
            {Ca-Ca} & {3.5+3.5 TeV} & $5.10^{+4.77}_{-2.53}$ pb & $3.83^{+4.71}_{-3.46}$ pb \\
            \hline
            {Ar-Ar} & {3.15+3.15 TeV} & $3.10^{+2.90}_{-1.54}$ pb & $2.33^{+2.87}_{-2.11}$ pb \\
            \hline
            {Kr-Kr} & {3.23+3.23 TeV} & $0.04^{+0.04}_{-0.02}$ nb & $0.03^{+0.04}_{-0.03}$ nb \\
            \hline
            {Xe-Xe} & {2.93+2.93 TeV}& $0.17^{+0.16}_{-0.09}$ nb & $0.13^{+0.16}_{-0.12}$ nb \\
            \hline
            {Pb-Pb} & {2.76+2.76 TeV} & $0.76^{+0.71}_{-0.38}$ nb & $0.57^{+0.70}_{-0.51}$ nb \\
            \hline
		\end{tabular}
    }
\end{table}

\begin{table}[h]
	\centering
    	\caption{Photon fusion cross sections of $T^{0++}_{4c}$ and $T^{2++}_{4c}$  for different collision center of mass energy $\sqrt{s}$ in electron-positron collision processes $(e^-\;e^+\,\xrightarrow{\gamma\gamma} e^-\;e^+\; T_{4c})$ at SCTF, Belle and CEPC with $\theta=\pi/4$ and ${\mathcal{B}(T_{4c})}=0.5$.}
        
	\label{tab:csee}
	\scalebox{0.95}{
		\begin{tabular}{|c|c|c|c|}
			\hline
            Experiments & $\sqrt{s}$ (GeV) &  $\sigma(T^{0++}_{4c})$ & $\sigma(T^{2++}_{4c})$ \\
            \hline
            {SCTF} & {7} & $0.18^{+0.17}_{-0.09}$ ab & $0.14^{+0.17}_{-0.13}$ ab \\
            \hline
            {Belle} & {10.6} & $5.25^{+4.90}_{-2.61}$ ab & $3.95^{+4.85}_{-3.56}$ ab \\
            \hline
            {CEPC} & {91.2} & $0.11^{+0.11}_{-0.06}$ fb & $0.08^{+0.10}_{-0.08}$ fb \\
            \hline
            {CEPC} & {172} & $0.17^{+0.16}_{-0.08}$ fb & $0.13^{+0.16}_{-0.12}$ fb \\
            \hline
            {CEPC} & {240} & $0.20^{+0.19}_{-0.10}$ fb & $0.15^{+0.19}_{-0.14}$ fb \\
            \hline
		\end{tabular}
    }
\end{table}

For the photon fusion in UPCs of proton-proton (P-P) and nucleus-nucleus (A-B), the electromagnetic field of the charged particles  accelerated at high energies can be identified as a flux of quasireal photons in the equivalent photon approximation. In this procedure we use $\theta=\pi/4$ and assume that ${\mathcal{B}(T_{4c})}=0.5$. We then give the 
photon fusion cross section in various UPC processes in Table \ref{tab:cs}. The UPC cross section of fully charm tetraquarks is not small. The cross section via photon fusion in proton-proton collision is around 0.1 pb, while the related cross section becomes around $\mu b$ in 
Lead-Lead collision because of the electromagnetic enhancing factor, i.e. the square of its electric charge $Z^2$. We also show our results of photon fusion cross sections of $T^{0++}_{4c}$ and $T^{2++}_{4c}$  for different collision center of mass energy $\sqrt{s}$ in electron-positron collision processes $(e^-\;e^+\,\xrightarrow{\gamma\gamma} e^-\;e^+\; T_{4c})$ in Table \ref{tab:csee}. From the table, the cross sections of fully charm tetraquarks are small. Nonetheless, these physical findings offer valuable guidance for the feasibility studies of future planned experimental facilities such as STCF and CEPC regarding fully charm tetraquark states~\cite{Ai:2025xop,Achasov:2023gey,CEPCStudyGroup:2023quu}.

\section{Conclusion}

In this paper, we investigate the interaction strength between fully charm tetraquark states and two photons. We obtain the next-to-leading order corrections to the decay widths of fully charm tetraquark states decaying into two photons. The clean two-photon channel will improve the signal recognition. Our final results, however, show that detecting this decay channel poses significant challenges in present experiments given that the partial decay width for the two-photon channel is on the order of $10^{-8}$ MeV, while the total width of fully charm tetraquark is around $100$ MeV. Nevertheless, we believe that future experiments at the LHC may make its observation possible. 

On the other hand, the production cross sections via photon fusion in various ultra-peripheral collisions and electron-positron collision processes are also investigated. Our results show that the photon fusion cross sections of $T^{0++}_{4c}$ and $T^{2++}_{4c}$ in various UPCs and electron-positron collision processes may be detected. Based on the cross section results for the fully charm tetraquarks, there are approximately 
3 signal events when using
 a data sample corresponding
to an integrated luminosity of 5 $\text{nb}^{-1}$ in Lead-Lead collision %and approximately \revise{1000} signal events when using a large data sample corresponding to an integrated luminosity of $1ab^{-1}$ in STCF experiments.
These results can be tested experimentally at the current LHC/Belle experiments and future STCF/CEPC experiments, and also provide a relatively clean approach for extracting the non-perturbative LDMEs of fully charm tetraquark states. Through studies of electromagnetic properties of  fully charm tetraquark states in current proton-proton or nucleus-nucleus collisions, a deeper understanding of their internal structure is anticipated.

\section*{Acknowledgments.}
%We thank the useful discussion with Prof. Xiao-Rong Zhou. 
This work is supported by
the National Natural Science Foundation of China
Grants No.12322503 and No.12405117.


\begin{thebibliography}{99}
%\cite{Choi:2003ue}
\bibitem{Choi:2003ue}
S.~K.~Choi \textit{et al.} [Belle],
%``Observation of a narrow charmonium-like state in exclusive $B^\pm \to K^\pm \pi^+ \pi^- J/\psi$ decays,''
Phys. Rev. Lett. \textbf{91}, 262001 (2003)
doi:10.1103/PhysRevLett.91.262001
[arXiv:hep-ex/0309032 [hep-ex]].
%2870 citations counted in INSPIRE as of 25 Dec 2025

%\cite{Acosta:2003zx}
\bibitem{Acosta:2003zx}
D.~Acosta \textit{et al.} [CDF],
%``Observation of the narrow state $X(3872) \to J/\psi \pi^+ \pi^-$ in $\bar{p}p$ collisions at $\sqrt{s} = 1.96$ TeV,''
Phys. Rev. Lett. \textbf{93}, 072001 (2004)
doi:10.1103/PhysRevLett.93.072001
[arXiv:hep-ex/0312021 [hep-ex]].
%1000 citations counted in INSPIRE as of 22 Dec 2025

%\cite{LHCb:2020bwg}
\bibitem{LHCb:2020bwg}
R.~Aaij \textit{et al.} [LHCb],
%``Observation of structure in the $J /\psi$ -pair mass spectrum,''
Sci. Bull. \textbf{65}, no.23, 1983-1993 (2020)
doi:10.1016/j.scib.2020.08.032
[arXiv:2006.16957 [hep-ex]].
%541 citations counted in INSPIRE as of 23 Dec 2025

%\cite{ATLAS:2023bft}
\bibitem{ATLAS:2023bft}
G.~Aad \textit{et al.} [ATLAS],
%``Observation of an Excess of Dicharmonium Events in the Four-Muon Final State with the ATLAS Detector,''
Phys. Rev. Lett. \textbf{131}, no.15, 151902 (2023)
doi:10.1103/PhysRevLett.131.151902
[arXiv:2304.08962 [hep-ex]].
%146 citations counted in INSPIRE as of 23 Dec 2025

%\cite{CMS:2023owd}
\bibitem{CMS:2023owd}
A.~Hayrapetyan \textit{et al.} [CMS],
%``New Structures in the J/{\ensuremath{\psi}}J/{\ensuremath{\psi}} Mass Spectrum in Proton-Proton Collisions at s=13{\,}{\,}TeV,''
Phys. Rev. Lett. \textbf{132}, no.11, 111901 (2024)
doi:10.1103/PhysRevLett.132.111901
[arXiv:2306.07164 [hep-ex]].
%187 citations counted in INSPIRE as of 23 Dec 2025

%\cite{Belle:2023gln}
\bibitem{Belle:2023gln}
J.~H.~Yin \textit{et al.} [Belle],
%``Search for the double-charmonium state with {\ensuremath{\eta}}$_{c}$J/{\ensuremath{\psi}} at Belle,''
JHEP \textbf{08}, 121 (2023)
doi:10.1007/JHEP08(2023)121
[arXiv:2305.17947 [hep-ex]].
%10 citations counted in INSPIRE as of 19 Dec 2025

%\cite{CMS:2025fpt}
\bibitem{CMS:2025fpt}
A.~Hayrapetyan \textit{et al.} [CMS],
%``Determination of the spin and parity of all-charm tetraquarks,''
Nature \textbf{648}, no.8092, 58-63 (2025)
doi:10.1038/s41586-025-09711-7
[arXiv:2506.07944 [hep-ex]].
%13 citations counted in INSPIRE as of 23 Dec 2025

%\cite{ATLAS:2025nsd}
\bibitem{ATLAS:2025nsd}
G.~Aad \textit{et al.} [ATLAS],
%``Observation of structures in the $J/ψ+ψ(2S)$ mass spectrum with the ATLAS detector,''
[arXiv:2509.13101 [hep-ex]].
%2 citations counted in INSPIRE as of 22 Dec 2025

%\cite{Iwasaki:1975pv}
\bibitem{Iwasaki:1975pv}
Y.~Iwasaki,
%``A Possible Model for New Resonances-Exotics and Hidden Charm,''
Prog. Theor. Phys. \textbf{54}, 492 (1975)
doi:10.1143/PTP.54.492
%100 citations counted in INSPIRE as of 04 Dec 2025

%\cite{Chao:1980dv}
\bibitem{Chao:1980dv}
K.~T.~Chao,
%``The (cc) - ($\bar{cc}$) (Diquark - Anti-Diquark) States in $e^+ e^-$ Annihilation,''
Z. Phys. C \textbf{7}, 317 (1981)
doi:10.1007/BF01431564
%106 citations counted in INSPIRE as of 04 Dec 2025

%\cite{Anwar:2017toa}
\bibitem{Anwar:2017toa}
M.~N.~Anwar, J.~Ferretti, F.~K.~Guo, E.~Santopinto and B.~S.~Zou,
%``Spectroscopy and decays of the fully-heavy tetraquarks,''
Eur. Phys. J. C \textbf{78}, no.8, 647 (2018)
doi:10.1140/epjc/s10052-018-6073-9
[arXiv:1710.02540 [hep-ph]].
%160 citations counted in INSPIRE as of 17 Dec 2025

%\cite{Karliner:2016zzc}
\bibitem{Karliner:2016zzc}
M.~Karliner, S.~Nussinov and J.~L.~Rosner,
%``$Q Q \bar Q \bar Q$ states: masses, production, and decays,''
Phys. Rev. D \textbf{95}, no.3, 034011 (2017)
doi:10.1103/PhysRevD.95.034011
[arXiv:1611.00348 [hep-ph]].
%200 citations counted in INSPIRE as of 10 Dec 2025

%\cite{Debastiani:2017msn}
\bibitem{Debastiani:2017msn}
V.~R.~Debastiani and F.~S.~Navarra,
%``A non-relativistic model for the $[cc][\bar{c}\bar{c}]$ tetraquark,''
Chin. Phys. C \textbf{43}, no.1, 013105 (2019)
doi:10.1088/1674-1137/43/1/013105
[arXiv:1706.07553 [hep-ph]].
%175 citations counted in INSPIRE as of 10 Dec 2025

%\cite{Wu:2016vtq}
\bibitem{Wu:2016vtq}
J.~Wu, Y.~R.~Liu, K.~Chen, X.~Liu and S.~L.~Zhu,
%``Heavy-flavored tetraquark states with the $QQ\bar{Q}\bar{Q}$ configuration,''
Phys. Rev. D \textbf{97}, no.9, 094015 (2018)
doi:10.1103/PhysRevD.97.094015
[arXiv:1605.01134 [hep-ph]].
%200 citations counted in INSPIRE as of 10 Dec 2025

%\cite{Liu:2019zuc}
\bibitem{Liu:2019zuc}
M.~S.~Liu, Q.~F.~L{\"u}, X.~H.~Zhong and Q.~Zhao,
%``All-heavy tetraquarks,''
Phys. Rev. D \textbf{100}, no.1, 016006 (2019)
doi:10.1103/PhysRevD.100.016006
[arXiv:1901.02564 [hep-ph]].
%159 citations counted in INSPIRE as of 10 Dec 2025

%\cite{Jin:2020jfc}
\bibitem{Jin:2020jfc}
X.~Jin, Y.~Xue, H.~Huang and J.~Ping,
%``Full-heavy tetraquarks in constituent quark models,''
Eur. Phys. J. C \textbf{80}, no.11, 1083 (2020)
doi:10.1140/epjc/s10052-020-08650-z
[arXiv:2006.13745 [hep-ph]].
%114 citations counted in INSPIRE as of 10 Dec 2025

%\cite{Liu:2021rtn}
\bibitem{Liu:2021rtn}
F.~X.~Liu, M.~S.~Liu, X.~H.~Zhong and Q.~Zhao,
%``Higher mass spectra of the fully-charmed and fully-bottom tetraquarks,''
Phys. Rev. D \textbf{104}, no.11, 116029 (2021)
doi:10.1103/PhysRevD.104.116029
[arXiv:2110.09052 [hep-ph]].
%55 citations counted in INSPIRE as of 10 Dec 2025

%\cite{Faustov:2020qfm}
\bibitem{Faustov:2020qfm}
R.~N.~Faustov, V.~O.~Galkin and E.~M.~Savchenko,
%``Masses of the $QQ\bar Q\bar Q$ tetraquarks in the relativistic diquark--antidiquark picture,''
Phys. Rev. D \textbf{102}, no.11, 114030 (2020)
doi:10.1103/PhysRevD.102.114030
[arXiv:2009.13237 [hep-ph]].
%83 citations counted in INSPIRE as of 04 Dec 2025

%\cite{Lu:2020cns}
\bibitem{Lu:2020cns}
Q.~F.~L{\"u}, D.~Y.~Chen and Y.~B.~Dong,
%``Masses of fully heavy tetraquarks $QQ {\bar{Q}} {\bar{Q}}$ in an extended relativized quark model,''
Eur. Phys. J. C \textbf{80}, no.9, 871 (2020)
doi:10.1140/epjc/s10052-020-08454-1
[arXiv:2006.14445 [hep-ph]].
%145 citations counted in INSPIRE as of 10 Dec 2025

%\cite{Liu:2020eha}
\bibitem{liu:2020eha}
M.~S.~Liu, F.~X.~Liu, X.~H.~Zhong and Q.~Zhao,
%``Fully heavy tetraquark states and their evidences in LHC observations,''
Phys. Rev. D \textbf{109}, no.7, 076017 (2024)
doi:10.1103/PhysRevD.109.076017
[arXiv:2006.11952 [hep-ph]].
%111 citations counted in INSPIRE as of 10 Dec 2025

%\cite{Wang:2019rdo}
\bibitem{Wang:2019rdo}
G.~J.~Wang, L.~Meng and S.~L.~Zhu,
%``Spectrum of the fully-heavy tetraquark state $QQ\bar Q' \bar Q'$,''
Phys. Rev. D \textbf{100}, no.9, 096013 (2019)
doi:10.1103/PhysRevD.100.096013
[arXiv:1907.05177 [hep-ph]].
%125 citations counted in INSPIRE as of 10 Dec 2025

%\cite{Bedolla:2019zwg}
\bibitem{Bedolla:2019zwg}
M.~A.~Bedolla, J.~Ferretti, C.~D.~Roberts and E.~Santopinto,
%``Spectrum of fully-heavy tetraquarks from a diquark+antidiquark perspective,''
Eur. Phys. J. C \textbf{80}, no.11, 1004 (2020)
doi:10.1140/epjc/s10052-020-08579-3
[arXiv:1911.00960 [hep-ph]].
%168 citations counted in INSPIRE as of 23 Dec 2025

%\cite{Zhu:2020xni}
\bibitem{Zhu:2020xni}
R.~Zhu,
%``Fully-heavy tetraquark spectra and production at hadron colliders,''
Nucl. Phys. B \textbf{966}, 115393 (2021)
doi:10.1016/j.nuclphysb.2021.115393
[arXiv:2010.09082 [hep-ph]].
%91 citations counted in INSPIRE as of 12 Dec 2025

%\cite{Giron:2020wpx}
\bibitem{Giron:2020wpx}
J.~F.~Giron and R.~F.~Lebed,
%``Simple spectrum of $c\bar c c\bar c$ states in the dynamical diquark model,''
Phys. Rev. D \textbf{102}, no.7, 074003 (2020)
doi:10.1103/PhysRevD.102.074003
[arXiv:2008.01631 [hep-ph]].
%120 citations counted in INSPIRE as of 04 Dec 2025

%\cite{Tang:2024zvf}
\bibitem{Tang:2024zvf}
C.~M.~Tang, C.~G.~Duan and L.~Tang,
%``full charmed tetraquark states in $8_{[c\bar{c}]}\otimes 8_{[c\bar{c}]}$ color structure via QCD sum rules,''
Eur. Phys. J. C \textbf{84}, no.7, 743 (2024)
doi:10.1140/epjc/s10052-024-13102-z
[arXiv:2405.05042 [hep-ph]].
%17 citations counted in INSPIRE as of 16 Dec 2025

%\cite{Tang:2024kmh}
\bibitem{Tang:2024kmh}
C.~M.~Tang, C.~G.~Duan, L.~Tang and C.~F.~Qiao,
%``A novel configuration of gluonic tetraquark state,''
Eur. Phys. J. C \textbf{85}, no.4, 396 (2025)
doi:10.1140/epjc/s10052-025-14106-z
[arXiv:2411.11433 [hep-ph]].
%9 citations counted in INSPIRE as of 04 Dec 2025

%\cite{Tang:2025ept}
\bibitem{Tang:2025ept}
C.~M.~Tang, C.~G.~Duan, L.~Tang and C.~F.~Qiao,
%``QCD sum rule predictions on gluonic tetraquark states with $J^{PC}=0^{+-},0^{--}$ and $1^{\pm \pm}$,''
[arXiv:2511.18807 [hep-ph]].
%0 citations counted in INSPIRE as of 04 Dec 2025

%\cite{Chen:2016jxd}
\bibitem{Chen:2016jxd}
W.~Chen, H.~X.~Chen, X.~Liu, T.~G.~Steele and S.~L.~Zhu,
%``Hunting for exotic doubly hidden-charm/bottom tetraquark states,''
Phys. Lett. B \textbf{773}, 247-251 (2017)
doi:10.1016/j.physletb.2017.08.034
[arXiv:1605.01647 [hep-ph]].
%183 citations counted in INSPIRE as of 23 Dec 2025

%\cite{Zhang:2020xtb}
\bibitem{Zhang:2020xtb}
J.~R.~Zhang,
%``$0^{+}$ fully-charmed tetraquark states,''
Phys. Rev. D \textbf{103}, no.1, 014018 (2021)
doi:10.1103/PhysRevD.103.014018
[arXiv:2010.07719 [hep-ph]].
%77 citations counted in INSPIRE as of 04 Dec 2025

%\cite{Wang:2020dlo}
\bibitem{Wang:2020dlo}
Z.~G.~Wang,
%``Revisit the tetraquark candidates in the $J/\psi J/\psi$ mass spectrum,''
Int. J. Mod. Phys. A \textbf{36}, no.02, 2150014 (2021)
doi:10.1142/S0217751X21500147
[arXiv:2009.05371 [hep-ph]].
%38 citations counted in INSPIRE as of 04 Dec 2025

%\cite{Wang:2020ols}
\bibitem{Wang:2020ols}
Z.~G.~Wang,
%``Tetraquark candidates in the LHCb's di-$J/\psi$ mass spectrum,''
Chin. Phys. C \textbf{44}, no.11, 113106 (2020)
doi:10.1088/1674-1137/abb080
[arXiv:2006.13028 [hep-ph]].
%79 citations counted in INSPIRE as of 04 Dec 2025

%\cite{Albuquerque:2020hio}
\bibitem{Albuquerque:2020hio}
R.~M.~Albuquerque, S.~Narison, A.~Rabemananjara, D.~Rabetiarivony and G.~Randriamanatrika,
%``Doubly-hidden scalar heavy molecules and tetraquarks states from QCD at NLO,''
Phys. Rev. D \textbf{102}, no.9, 094001 (2020)
doi:10.1103/PhysRevD.102.094001
[arXiv:2008.01569 [hep-ph]].
%97 citations counted in INSPIRE as of 04 Dec 2025

%\cite{Wang:2021mma}
\bibitem{Wang:2021mma}
Q.~N.~Wang, Z.~Y.~Yang and W.~Chen,
%``Exotic fully-heavy $Q\bar QQ\bar Q$ tetraquark states in $\mathbf{8}_{[Q\bar{Q}]}\otimes \mathbf{8}_{[Q\bar{Q}]}$ color configuration,''
Phys. Rev. D \textbf{104}, no.11, 114037 (2021)
doi:10.1103/PhysRevD.104.114037
[arXiv:2109.08091 [hep-ph]].
%29 citations counted in INSPIRE as of 17 Dec 2025

%\cite{Wan:2020fsk}
\bibitem{Wan:2020fsk}
B.~D.~Wan and C.~F.~Qiao,
%``Gluonic tetracharm configuration of $X (6900)$,''
Phys. Lett. B \textbf{817}, 136339 (2021)
doi:10.1016/j.physletb.2021.136339
[arXiv:2012.00454 [hep-ph]].
%86 citations counted in INSPIRE as of 16 Dec 2025

%\cite{Yang:2020wkh}
\bibitem{Yang:2020wkh}
B.~C.~Yang, L.~Tang and C.~F.~Qiao,
%``Scalar fully-heavy tetraquark states $QQ^\prime {\bar{Q}} \bar{Q^\prime }$ in QCD sum rules,''
Eur. Phys. J. C \textbf{81}, no.4, 324 (2021)
doi:10.1140/epjc/s10052-021-09096-7
[arXiv:2012.04463 [hep-ph]].
%83 citations counted in INSPIRE as of 19 Dec 2025

%\cite{Wang:2017jtz}
\bibitem{Wang:2017jtz}
Z.~G.~Wang,
%``Analysis of the $QQ\bar{Q}\bar{Q}$ tetraquark states with QCD sum rules,''
Eur. Phys. J. C \textbf{77}, no.7, 432 (2017)
doi:10.1140/epjc/s10052-017-4997-0
[arXiv:1701.04285 [hep-ph]].
%140 citations counted in INSPIRE as of 08 Dec 2025

%\cite{Wang:2018poa}
\bibitem{Wang:2018poa}
Z.~G.~Wang and Z.~Y.~Di,
%``Analysis of the vector and axialvector $QQ\bar{Q}\bar{Q}$ tetraquark states with QCD sum rules,''
Acta Phys. Polon. B \textbf{50}, 1335 (2019)
doi:10.5506/APhysPolB.50.1335
[arXiv:1807.08520 [hep-ph]].
%78 citations counted in INSPIRE as of 08 Dec 2025

%\cite{Agaev:2023wua}
\bibitem{Agaev:2023wua}
S.~S.~Agaev, K.~Azizi, B.~Barsbay and H.~Sundu,
%``Exploring fully heavy scalar tetraquarks QQQ‾Q‾,''
Phys. Lett. B \textbf{844}, 138089 (2023)
doi:10.1016/j.physletb.2023.138089
[arXiv:2304.03244 [hep-ph]].
%35 citations counted in INSPIRE as of 19 Dec 2025

%\cite{Agaev:2023gaq}
\bibitem{Agaev:2023gaq}
S.~S.~Agaev, K.~Azizi, B.~Barsbay and H.~Sundu,
%``fully charm resonance X(6900) and its beauty counterpart,''
Nucl. Phys. A \textbf{1041}, 122768 (2024)
doi:10.1016/j.nuclphysa.2023.122768
[arXiv:2304.09943 [hep-ph]].
%24 citations counted in INSPIRE as of 19 Dec 2025

%\cite{Li:2025vbd}
\bibitem{Li:2025vbd}
G.~Li, C.~Shi, Y.~Chen and W.~Sun,
%``$\eta_c\eta_c$ and $J/\psi J/\psi$ scatterings from lattice QCD,''
[arXiv:2505.23220 [hep-lat]].
%2 citations counted in INSPIRE as of 04 Dec 2025

%\cite{Li:2025ftn}
\bibitem{Li:2025ftn}
G.~Li, C.~Shi, Y.~Chen and W.~Sun,
%``Tensor Resonance in $J/\psi J/\psi$ Scattering from Lattice QCD,''
[arXiv:2505.24213 [hep-lat]].
%1 citations counted in INSPIRE as of 04 Dec 2025

%\cite{Meng:2024czd}
\bibitem{Meng:2024czd}
Y.~Meng, C.~Liu, X.~Y.~Tuo, H.~Yan and Z.~Zhang,
%``Lattice calculation of the $\eta _c\eta _c$ and $J/\psi J/\psi $ s-wave scattering length,''
Eur. Phys. J. C \textbf{85}, no.4, 458 (2025)
doi:10.1140/epjc/s10052-025-14192-z
[arXiv:2411.11533 [hep-lat]].
%4 citations counted in INSPIRE as of 04 Dec 2025

%\cite{Santowsky:2021bhy}
\bibitem{Santowsky:2021bhy}
N.~Santowsky and C.~S.~Fischer,
%``Four-quark states with charm quarks in a two-body Bethe{\textendash}Salpeter approach,''
Eur. Phys. J. C \textbf{82}, no.4, 313 (2022)
doi:10.1140/epjc/s10052-022-10272-6
[arXiv:2111.15310 [hep-ph]].
%37 citations counted in INSPIRE as of 04 Dec 2025

%\cite{Li:2021ygk}
\bibitem{Li:2021ygk}
Q.~Li, C.~H.~Chang, G.~L.~Wang and T.~Wang,
%``Mass spectra and wave functions of TQQQ{\textasciimacron}Q{\textasciimacron} tetraquarks,''
Phys. Rev. D \textbf{104}, no.1, 014018 (2021)
doi:10.1103/PhysRevD.104.014018
[arXiv:2104.12372 [hep-ph]].
%53 citations counted in INSPIRE as of 04 Dec 2025

%\cite{Ke:2021iyh}
\bibitem{Ke:2021iyh}
H.~W.~Ke, X.~Han, X.~H.~Liu and Y.~L.~Shi,
%``Tetraquark state $X(6900)$ and the interaction between diquark and antidiquark,''
Eur. Phys. J. C \textbf{81}, no.5, 427 (2021)
doi:10.1140/epjc/s10052-021-09229-y
[arXiv:2103.13140 [hep-ph]].
%61 citations counted in INSPIRE as of 15 Dec 2025

%\cite{Heupel:2012ua}
\bibitem{Heupel:2012ua}
W.~Heupel, G.~Eichmann and C.~S.~Fischer,
%``Tetraquark Bound States in a Bethe-Salpeter Approach,''
Phys. Lett. B \textbf{718}, 545-549 (2012)
doi:10.1016/j.physletb.2012.11.009
[arXiv:1206.5129 [hep-ph]].
%136 citations counted in INSPIRE as of 16 Dec 2025

%\cite{Wang:2020wrp}
\bibitem{Wang:2020wrp}
J.~Z.~Wang, D.~Y.~Chen, X.~Liu and T.~Matsuki,
%``Producing fully charm structures in the $J/\psi$ -pair invariant mass spectrum,''
Phys. Rev. D \textbf{103}, no.7, 071503 (2021)
doi:10.1103/PhysRevD.103.L071503
[arXiv:2008.07430 [hep-ph]].
%75 citations counted in INSPIRE as of 10 Dec 2025

%\cite{Gong:2020bmg}
\bibitem{Gong:2020bmg}
C.~Gong, M.~C.~Du, Q.~Zhao, X.~H.~Zhong and B.~Zhou,
%``Nature of X(6900) and its production mechanism at LHCb,''
Phys. Lett. B \textbf{824}, 136794 (2022)
doi:10.1016/j.physletb.2021.136794
[arXiv:2011.11374 [hep-ph]].
%73 citations counted in INSPIRE as of 04 Dec 2025

%\cite{Dong:2020nwy}
\bibitem{Dong:2020nwy}
X.~K.~Dong, V.~Baru, F.~K.~Guo, C.~Hanhart and A.~Nefediev,
%``Coupled-Channel Interpretation of the LHCb Double-~$J/\psi$~Spectrum and Hints of a New State Near the~ $J/\psi J/\psi$~~Threshold,''
Phys. Rev. Lett. \textbf{126}, no.13, 132001 (2021)
[erratum: Phys. Rev. Lett. \textbf{127}, no.11, 119901 (2021)]
doi:10.1103/PhysRevLett.127.119901
[arXiv:2009.07795 [hep-ph]].
%133 citations counted in INSPIRE as of 12 Dec 2025

%\cite{Guo:2020pvt}
\bibitem{Guo:2020pvt}
Z.~H.~Guo and J.~A.~Oller,
%``Insights into the inner structures of the fully charm tetraquark state $X(6900)$,''
Phys. Rev. D \textbf{103}, no.3, 034024 (2021)
doi:10.1103/PhysRevD.103.034024
[arXiv:2011.00978 [hep-ph]].
%71 citations counted in INSPIRE as of 04 Dec 2025

%\cite{Huang:2024jin}
\bibitem{Huang:2024jin}
Q.~Huang, R.~Chen, J.~He and X.~Liu,
%``Discovering a Novel Dynamics Mechanism for Charmonium Scattering,''
[arXiv:2407.16316 [hep-ph]].
%8 citations counted in INSPIRE as of 19 Dec 2025

%\cite{Esposito:2018cwh}
\bibitem{Esposito:2018cwh}
A.~Esposito and A.~D.~Polosa,
%``A $bb\bar b\bar b$ di-bottomonium at the LHC?,''
Eur. Phys. J. C \textbf{78}, no.9, 782 (2018)
doi:10.1140/epjc/s10052-018-6269-z
[arXiv:1807.06040 [hep-ph]].
%106 citations counted in INSPIRE as of 20 Dec 2025

%\cite{Agaev:2025wyf}
\bibitem{Agaev:2025wyf}
S.~S.~Agaev, K.~Azizi and H.~Sundu,
%``Axial-vector molecules $ΥB_{c}^{-}$ and $η_{b}B_{c}^{\ast-} $,''
[arXiv:2512.06513 [hep-ph]].
%0 citations counted in INSPIRE as of 10 Dec 2025

%\cite{Wang:2025apq}
\bibitem{Wang:2025apq}
Z.~Y.~Wang, J.~J.~Qi, Z.~H.~Zhang and X.~H.~Guo,
%``Spectra of bcb{\textasciimacron}c{\textasciimacron} tetraquark states from a diquark-antidiquark perspective,''
Phys. Rev. D \textbf{112}, no.7, 074038 (2025)
doi:10.1103/skdp-g4ql
[arXiv:2508.12366 [hep-ph]].
%0 citations counted in INSPIRE as of 04 Dec 2025

%\cite{Xia:2025mgk}
\bibitem{Xia:2025mgk}
X.~Xia and T.~Guo,
%``Fully heavy tetraquark states with diquark-antidiquark configuration,''
[arXiv:2508.19838 [hep-ph]].
%0 citations counted in INSPIRE as of 04 Dec 2025

%\cite{Feng:2023agq}
\bibitem{Feng:2023agq}
F.~Feng, Y.~Huang, Y.~Jia, W.~L.~Sang, D.~S.~Yang and J.~Y.~Zhang,
%``Inclusive production of fully charm tetraquarks at the LHC,''
Phys. Rev. D \textbf{108}, no.5, L051501 (2023)
doi:10.1103/PhysRevD.108.L051501
[arXiv:2304.11142 [hep-ph]].
%28 citations counted in INSPIRE as of 10 Dec 2025

%\cite{Belov:2024qyi}
\bibitem{Belov:2024qyi}
I.~Belov, A.~Giachino and E.~Santopinto,
%``fully charm tetraquark production at the LHC experiments,''
JHEP \textbf{01}, 093 (2025)
doi:10.1007/JHEP01(2025)093
[arXiv:2409.12070 [hep-ph]].
%8 citations counted in INSPIRE as of 10 Dec 2025

%\cite{Wang:2025hex}
\bibitem{Wang:2025hex}
Y.~Wang and R.~Zhu,
%``fully charm tetraquark production at hadronic collisions with gluon radiation effects,''
[arXiv:2510.02085 [hep-ph]].
%1 citations counted in INSPIRE as of 19 Dec 2025

%\cite{Celiberto:2025vra}
\bibitem{Celiberto:2025vra}
F.~G.~Celiberto, A.~V.~Giannini, V.~P.~Gon{\c{c}}alves and Y.~N.~Lima,
%``fully charm tetraquark production in forward rapidity $pp$ collisions at LHC and FCC energies,''
[arXiv:2511.18984 [hep-ph]].
%1 citations counted in INSPIRE as of 20 Dec 2025

%\cite{Celiberto:2025ziy}
\bibitem{Celiberto:2025ziy}
F.~G.~Celiberto,
%``Fragmentation of fully heavy tetraquarks: The TQ4Q1.1 functions as a case study,''
Phys. Rev. D \textbf{112}, no.7, 074041 (2025)
doi:10.1103/375n-fw5h
[arXiv:2507.09744 [hep-ph]].
%10 citations counted in INSPIRE as of 19 Dec 2025

%\cite{Zhang:2020hoh}
\bibitem{Zhang:2020hoh}
H.~F.~Zhang, Y.~Q.~Ma and W.~L.~Sang,
%``Perturbative QCD evidence for spin-2 particles in the di-J/{\ensuremath{\psi}} resonances,''
Sci. Bull. \textbf{70}, 1915-1917 (2025)
doi:10.1016/j.scib.2025.04.035
[arXiv:2009.08376 [hep-ph]].
%70 citations counted in INSPIRE as of 10 Dec 2025

%\cite{Feng:2020riv}
\bibitem{Feng:2020riv}
F.~Feng, Y.~Huang, Y.~Jia, W.~L.~Sang, X.~Xiong and J.~Y.~Zhang,
%``Fragmentation production of fully-charmed tetraquarks at the LHC,''
Phys. Rev. D \textbf{106}, no.11, 114029 (2022)
doi:10.1103/PhysRevD.106.114029
[arXiv:2009.08450 [hep-ph]].
%67 citations counted in INSPIRE as of 08 Dec 2025

%\cite{Bai:2024flh}
\bibitem{Bai:2024flh}
X.~W.~Bai, Y.~Huang and W.~L.~Sang,
%``Light quark fragmentation into an S-wave fully charm tetraquark,''
Phys. Rev. D \textbf{111}, no.5, 054006 (2025)
doi:10.1103/PhysRevD.111.054006
[arXiv:2411.19296 [hep-ph]].
%6 citations counted in INSPIRE as of 19 Dec 2025

%\cite{Bai:2024ezn}
\bibitem{Bai:2024ezn}
X.~W.~Bai, F.~Feng, C.~M.~Gan, Y.~Huang, W.~L.~Sang and H.~F.~Zhang,
%``Producing fully-charmed tetraquarks via charm quark fragmentation in colliders,''
JHEP \textbf{09}, 002 (2024)
doi:10.1007/JHEP09(2024)002
[arXiv:2404.13889 [hep-ph]].
%12 citations counted in INSPIRE as of 04 Dec 2025

%\cite{Celiberto:2024mab}
\bibitem{Celiberto:2024mab}
F.~G.~Celiberto, G.~Gatto and A.~Papa,
%``fully charm tetraquarks from LHC to FCC: natural stability from fragmentation,''
Eur. Phys. J. C \textbf{84}, no.10, 1071 (2024)
doi:10.1140/epjc/s10052-024-13345-w
[arXiv:2405.14773 [hep-ph]].
%23 citations counted in INSPIRE as of 19 Dec 2025

%\cite{Sang:2023ncm}
\bibitem{Sang:2023ncm}
W.~L.~Sang, T.~Wang, Y.~D.~Zhang and F.~Feng,
%``Electromagnetic and hadronic decay of fully heavy tetraquarks,''
Phys. Rev. D \textbf{109}, no.5, 056016 (2024)
doi:10.1103/PhysRevD.109.056016
[arXiv:2307.16150 [hep-ph]].
%16 citations counted in INSPIRE as of 23 Dec 2025

%\cite{Chen:2024orv}
\bibitem{Chen:2024orv}
K.~Chen, F.~X.~Liu, Q.~Zhao, X.~H.~Zhong, R.~Zhu and B.~S.~Zou,
%``Decoding spin-parity quantum numbers and decay widths of double $J/ψ$ exotic states,''
[arXiv:2412.13455 [hep-ph]].
%1 citations counted in INSPIRE as of 19 Dec 2025

%\cite{Becchi:2020uvq}
\bibitem{Becchi:2020uvq}
C.~Becchi, J.~Ferretti, A.~Giachino, L.~Maiani and E.~Santopinto,
%``A study of $c c\bar{c}\bar{c}$ tetraquark decays in 4 muons and in $D^{(*)} \bar{D}^{(*)}$ at LHC,''
Phys. Lett. B \textbf{811}, 135952 (2020)
doi:10.1016/j.physletb.2020.135952
[arXiv:2006.14388 [hep-ph]].
%93 citations counted in INSPIRE as of 10 Dec 2025

%\cite{Zhang:2023ffe}
\bibitem{Zhang:2023ffe}
H.~F.~Zhang, X.~M.~Mo and Y.~P.~Yan,
%``Hadronic decay of exotic mesons consisting of four charm quarks,''
Phys. Rev. D \textbf{110}, no.9, 096021 (2024)
doi:10.1103/PhysRevD.110.096021
[arXiv:2312.10850 [hep-ph]].
%5 citations counted in INSPIRE as of 04 Dec 2025

%\cite{Wang:2023kir}
\bibitem{Wang:2023kir}
Z.~G.~Wang and X.~S.~Yang,
%``The two-body strong decays of the fully charm tetraquark states,''
AAPPS Bull. \textbf{34}, 5 (2024)
doi:10.1007/s43673-023-00112-4
[arXiv:2310.16583 [hep-ph]].
%16 citations counted in INSPIRE as of 04 Dec 2025

%\cite{Biloshytskyi:2022dmo}
\bibitem{Biloshytskyi:2022dmo}
V.~Biloshytskyi, V.~Pascalutsa, L.~Harland-Lang, B.~Malaescu, K.~Schmieden and M.~Schott,
%``Two-photon decay of X(6900) from light-by-light scattering at the LHC,''
Phys. Rev. D \textbf{106}, no.11, L111902 (2022)
doi:10.1103/PhysRevD.106.L111902
[arXiv:2207.13623 [hep-ph]].
%17 citations counted in INSPIRE as of 17 Dec 2025

%\cite{Chen:2022sbf}
\bibitem{Chen:2022sbf}
H.~X.~Chen, Y.~X.~Yan and W.~Chen,
%``Decay behaviors of the fully bottom and fully charm tetraquark states,''
Phys. Rev. D \textbf{106}, no.9, 094019 (2022)
doi:10.1103/PhysRevD.106.094019
[arXiv:2207.08593 [hep-ph]].
%25 citations counted in INSPIRE as of 23 Dec 2025

%\cite{Lu:2025lyu}
\bibitem{Lu:2025lyu}
D.~D.~Lu and S.~Z.~Jiang,
%``Investigating the internal structure of $X(6900)$ in the $2J/ψ$ decay channel,''
[arXiv:2512.18569 [hep-ph]].
%0 citations counted in INSPIRE as of 24 Dec 2025

%\cite{Zhu:2024swp}
\bibitem{Zhu:2024swp}
F.~Zhu, G.~Bauer and K.~Yi,
%``Experimental Road to a Charming Family of Tetraquarks {\textellipsis} and Beyond,''
Chin. Phys. Lett. \textbf{41}, no.11, 111201 (2024)
doi:10.1088/0256-307X/41/11/111201
[arXiv:2410.11210 [hep-ph]].
%15 citations counted in INSPIRE as of 25 Dec 2025

%\cite{Bodwin:1994jh}
\bibitem{Bodwin:1994jh}
G.~T.~Bodwin, E.~Braaten and G.~P.~Lepage,
%``Rigorous QCD analysis of inclusive annihilation and production of heavy quarkonium,''
Phys. Rev. D \textbf{51}, 1125-1171 (1995)
[erratum: Phys. Rev. D \textbf{55}, 5853 (1997)]
doi:10.1103/PhysRevD.55.5853
[arXiv:hep-ph/9407339 [hep-ph]].
%3119 citations counted in INSPIRE as of 24 Dec 2025

%\cite{Zhu:2015bba}
\bibitem{Zhu:2015bba}
R.~Zhu and C.~F.~Qiao,
%``Pentaquark states in a diquark{\textendash}triquark model,''
Phys. Lett. B \textbf{756}, 259-264 (2016)
doi:10.1016/j.physletb.2016.03.022
[arXiv:1510.08693 [hep-ph]].
%156 citations counted in INSPIRE as of 04 Dec 2025

%\cite{Wang:2017vnc}
\bibitem{Wang:2017vnc}
W.~Wang and R.~L.~Zhu,
%``Interpretation of the newly observed $\Omega_c^0$ resonances,''
Phys. Rev. D \textbf{96}, no.1, 014024 (2017)
doi:10.1103/PhysRevD.96.014024
[arXiv:1704.00179 [hep-ph]].
%91 citations counted in INSPIRE as of 17 Dec 2025

%\cite{Zhu:2016arf}
\bibitem{Zhu:2016arf}
R.~Zhu,
%``Hidden charm octet tetraquarks from a diquark-antidiquark model,''
Phys. Rev. D \textbf{94}, no.5, 054009 (2016)
doi:10.1103/PhysRevD.94.054009
[arXiv:1607.02799 [hep-ph]].
%48 citations counted in INSPIRE as of 07 Dec 2025

%\cite{Qiao:2012hp}
\bibitem{Qiao:2012hp}
C.~F.~Qiao, P.~Sun, D.~Yang and R.~L.~Zhu,
%``B$_c$ exclusive decays to charmonium and a light meson at next-to-leading order accuracy,''
Phys. Rev. D \textbf{89}, no.3, 034008 (2014)
doi:10.1103/PhysRevD.89.034008
[arXiv:1209.5859 [hep-ph]].
%83 citations counted in INSPIRE as of 11 Dec 2025

%\cite{Qiao:2012vt}
\bibitem{Qiao:2012vt}
C.~F.~Qiao and R.~L.~Zhu,
%``Estimation of semileptonic decays of $B_c$ meson to S-wave charmonia with nonrelativistic QCD,''
Phys. Rev. D \textbf{87}, no.1, 014009 (2013)
doi:10.1103/PhysRevD.87.014009
[arXiv:1208.5916 [hep-ph]].
%64 citations counted in INSPIRE as of 04 Dec 2025

%\cite{Hahn:2000kx}
\bibitem{Hahn:2000kx}
T.~Hahn,
%``Generating Feynman diagrams and amplitudes with FeynArts 3,''
Comput. Phys. Commun. \textbf{140}, 418-431 (2001)
doi:10.1016/S0010-4655(01)00290-9
[arXiv:hep-ph/0012260 [hep-ph]].
%2392 citations counted in INSPIRE as of 24 Dec 2025

%\cite{Shtabovenko:2020gxv}
\bibitem{Shtabovenko:2020gxv}
V.~Shtabovenko, R.~Mertig and F.~Orellana,
%``FeynCalc 9.3: New features and improvements,''
Comput. Phys. Commun. \textbf{256}, 107478 (2020)
doi:10.1016/j.cpc.2020.107478
[arXiv:2001.04407 [hep-ph]].
%701 citations counted in INSPIRE as of 24 Dec 2025

%\cite{Shtabovenko:2023idz}
\bibitem{Shtabovenko:2023idz}
V.~Shtabovenko, R.~Mertig and F.~Orellana,
%``FeynCalc 10: Do multiloop integrals dream of computer codes?,''
Comput. Phys. Commun. \textbf{306}, 109357 (2025)
doi:10.1016/j.cpc.2024.109357
[arXiv:2312.14089 [hep-ph]].
%137 citations counted in INSPIRE as of 24 Dec 2025

%\cite{Tkachov:1981wb}
\bibitem{Tkachov:1981wb}
F.~V.~Tkachov,
%``A theorem on analytical calculability of 4-loop renormalization group functions,''
Phys. Lett. B \textbf{100}, 65-68 (1981)
doi:10.1016/0370-2693(81)90288-4
%1422 citations counted in INSPIRE as of 24 Dec 2025

%\cite{Chetyrkin:1981qh}
\bibitem{Chetyrkin:1981qh}
K.~G.~Chetyrkin and F.~V.~Tkachov,
%``Integration by parts: The algorithm to calculate $\beta$-functions in 4 loops,''
Nucl. Phys. B \textbf{192}, 159-204 (1981)
doi:10.1016/0550-3213(81)90199-1
%2364 citations counted in INSPIRE as of 25 Dec 2025

%\cite{Klappert:2020nbg}
\bibitem{Klappert:2020nbg}
J.~Klappert, F.~Lange, P.~Maierh{\"o}fer and J.~Usovitsch,
%``Integral reduction with Kira 2.0 and finite field methods,''
Comput. Phys. Commun. \textbf{266}, 108024 (2021)
doi:10.1016/j.cpc.2021.108024
[arXiv:2008.06494 [hep-ph]].
%400 citations counted in INSPIRE as of 24 Dec 2025

%\cite{Shtabovenko:2016whf}
\bibitem{Shtabovenko:2016whf}
V.~Shtabovenko,
%``FeynHelpers: Connecting FeynCalc to FIRE and Package-X,''
Comput. Phys. Commun. \textbf{218}, 48-65 (2017)
doi:10.1016/j.cpc.2017.04.014
[arXiv:1611.06793 [physics.comp-ph]].
%139 citations counted in INSPIRE as of 17 Dec 2025

%\cite{Patel:2015tea}
\bibitem{Patel:2015tea}
H.~H.~Patel,
%``Package-X: A Mathematica package for the analytic calculation of one-loop integrals,''
Comput. Phys. Commun. \textbf{197}, 276-290 (2015)
doi:10.1016/j.cpc.2015.08.017
[arXiv:1503.01469 [hep-ph]].
%597 citations counted in INSPIRE as of 24 Dec 2025

%\cite{Shao:2022cly}
\bibitem{Shao:2022cly}
H.~S.~Shao and D.~d'Enterria,
%``gamma-UPC: automated generation of exclusive photon-photon processes in ultraperipheral proton and nuclear collisions with varying form factors,''
JHEP \textbf{09}, 248 (2022)
doi:10.1007/JHEP09(2022)248
[arXiv:2207.03012 [hep-ph]].
%78 citations counted in INSPIRE as of 11 Dec 2025

%\cite{Jiang:2025pad}
\bibitem{Jiang:2025pad}
J.~Jiang, H.~Yang, X.~Liang, Z.~G.~Si, C.~F.~Qiao, B.~W.~Long, Y.~R.~Liu and S.~Y.~Li,
%``Doubly heavy hadron production in ultraperipheral collisions,''
[arXiv:2509.06258 [hep-ph]].
%2 citations counted in INSPIRE as of 04 Dec 2025

%\cite{ParticleDataGroup:2024cfk}
\bibitem{ParticleDataGroup:2024cfk}
S.~Navas \textit{et al.} [Particle Data Group],
%``Review of particle physics,''
Phys. Rev. D \textbf{110}, no.3, 030001 (2024)
doi:10.1103/PhysRevD.110.030001
%3539 citations counted in INSPIRE as of 26 Dec 2025


\bibitem{Ai:2025xop}
X.~C.~Ai, J.~Bao, L.~P.~An, S.~Z.~An, Z.~Cao, Y.~Bai, M.~Chang, Z.~H.~Bai, F.~Chen and O.~Bakina, \textit{et al.}
%``Conceptual design report of the Super Tau-Charm Facility: the accelerator,''
Nucl. Sci. Tech. \textbf{36}, no.12, 242 (2025)
doi:10.1007/s41365-025-01833-x
[arXiv:2509.11522 [physics.acc-ph]].
%4 citations counted in INSPIRE as of 26 Dec 2025

%\cite{Achasov:2023gey}
\bibitem{Achasov:2023gey}
M.~Achasov, X.~C.~Ai, R.~Aliberti, L.~P.~An, Q.~An, X.~Z.~Bai, Y.~Bai, O.~Bakina, A.~Barnyakov and V.~Blinov, \textit{et al.}
%``STCF conceptual design report (Volume 1): Physics {\&} detector,''
Front. Phys. (Beijing) \textbf{19}, no.1, 14701 (2024)
doi:10.1007/s11467-023-1333-z
[arXiv:2303.15790 [hep-ex]].
%203 citations counted in INSPIRE as of 26 Dec 2025


%\cite{CEPCStudyGroup:2023quu}
\bibitem{CEPCStudyGroup:2023quu}
W.~Abdallah \textit{et al.} [CEPC Study Group],
%``CEPC Technical Design Report: Accelerator,''
Radiat. Detect. Technol. Methods \textbf{8}, no.1, 1-1105 (2024)
[erratum: Radiat. Detect. Technol. Methods \textbf{9}, no.1, 184-192 (2025)]
doi:10.1007/s41605-024-00463-y
[arXiv:2312.14363 [physics.acc-ph]].
%201 citations counted in INSPIRE as of 26 Dec 2025
%\cite{Esposito:2021ptx}
\bibitem{Esposito:2021ptx}
A.~Esposito, C.~A.~Manzari, A.~Pilloni and A.~D.~Polosa,
%``Hunting for tetraquarks in ultraperipheral heavy ion collisions,''
Phys. Rev. D \textbf{104}, no.11, 114029 (2021)
doi:10.1103/PhysRevD.104.114029
[arXiv:2109.10359 [hep-ph]].
%32 citations counted in INSPIRE as of 30 May 2026
%\cite{dEnterria:2025ecx}
\bibitem{dEnterria:2025ecx}
D.~d'Enterria and K.~Kang,
%``Exclusive photon-fusion production of even-spin resonances and exotic QED atoms in high-energy hadron collisions,''
Phys. Rev. D \textbf{112}, no.11, 116022 (2025)
doi:10.1103/rnxl-v6gd
[arXiv:2503.10952 [hep-ph]].
%13 citations counted in INSPIRE as of 30 May 2026
%\cite{ParticleDataGroup:2024cfk}
\bibitem{ParticleDataGroup:2024cfk}
S.~Navas \textit{et al.} [Particle Data Group],
%``Review of particle physics,''
Phys. Rev. D \textbf{110}, no.3, 030001 (2024)
doi:10.1103/PhysRevD.110.030001
%5127 citations counted in INSPIRE as of 31 May 2026
%\cite{Herren:2017osy}
\bibitem{Herren:2017osy}
F.~Herren and M.~Steinhauser,
%``Version 3 of RunDec and CRunDec,''
Comput. Phys. Commun. \textbf{224}, 333-345 (2018)
doi:10.1016/j.cpc.2017.11.014
[arXiv:1703.03751 [hep-ph]].
%334 citations counted in INSPIRE as of 31 May 2026
\end{thebibliography}
\end{document}